\documentclass[12pt]{article}
\usepackage{a4wide}
\usepackage{bbold}
\usepackage{amsmath,amssymb,bbm}
\usepackage{multirow}
\usepackage{comment}
\newsavebox{\uuunit}
\sbox{\uuunit}
    {\setlength{\unitlength}{0.825em}
     \begin{picture}(0.6,0.7)
        \thinlines
        \put(0,0){\line(1,0){0.5}}
        \put(0.15,0){\line(0,1){0.7}}
        \put(0.35,0){\line(0,1){0.8}}
       \multiput(0.3,0.8)(-0.04,-0.02){12}{\rule{0.5pt}{0.5pt}}
     \end {picture}}

\def\2{\frac12}
\def\4{\frac14}

\newcommand{\be}{\begin{equation}}
\newcommand{\ee}{\end{equation}}
\newcommand{\bea}{\begin{eqnarray}}
\newcommand{\eea}{\end{eqnarray}}

\begin{document}

\begin{titlepage}%1
\begin{center}

\hfill UG-12-02 \\ \hfill CERN-PH-TH/2012-001

\vskip 1.5cm

{\LARGE \bf  Brane Orbits}

\vskip 1cm

{\bf Eric A.~Bergshoeff\,$^1$, Alessio Marrani\,$^2$  and Fabio Riccioni\,$^3$}

\vskip 25pt

{\em $^1$ \hskip -.1truecm Centre for Theoretical Physics,
University of Groningen, \\ Nijenborgh 4, 9747 AG Groningen, The
Netherlands \vskip 5pt }

{email: {\tt E.A.Bergshoeff@rug.nl}} \\

\vskip 15pt

{\em $^2$ \hskip -.1truecm Physics Department, Theory Unit, CERN,
CH -1211, Geneva 23, Switzerland  \vskip 5pt }

{email: {\tt  Alessio.Marrani@cern.ch}} \\

\vskip 15pt

{\em $^3$ \hskip -.1truecm
 INFN Sezione di Roma,  Dipartimento di Fisica, Universit\`a di Roma ``La Sapienza'',\\ Piazzale Aldo Moro 2, 00185 Roma, Italy
 \vskip 5pt }

{email: {\tt Fabio.Riccioni@roma1.infn.it}} \\

\end{center}

\vskip 0.5cm

\begin{center} {\bf ABSTRACT}\\[3ex]
\end{center}

We complete the classification of half-supersymmetric branes in
toroidally compactified IIA/IIB string theory in terms of
representations of the T-duality group. As a by-product we derive a
last wrapping rule for the space-filling branes. We find
examples of T-duality representations of branes in lower dimensions,
suggested by supergravity,
of which {\sl none} of the component branes follow from the
reduction of any brane in ten-dimensional IIA/IIB string theory. We
discuss the constraints on the charges of half-supersymmetric
branes, determining the corresponding T-duality and  U-duality
orbits.

\end{titlepage}

\newpage
\setcounter{page}{1} \tableofcontents

\newpage

%%%%%%%%%%%%%%%%%%%%%%%%%%%%%%%%%%%% %%%%%%%%%%%%%%%%%%%%%%%%%%%%%%%%%%%%%
%%%%%%%%%%%%%%%%%%%%
\setcounter{page}{1} \numberwithin{equation}{section}

\section{Introduction}

It is by now well-understood  that branes form a crucial ingredient
of string theory. For instance, they have been used to calculate the
entropy of certain black holes \cite{Strominger:1996sh} and they are
at the heart of the AdS/CFT correspondence \cite{Maldacena:1997re}.
In particular, the half-supersymmetric branes, i.e.~the ones with 16 supercharges,
play a relevant role in all these applications. It is therefore of great interest to
find out the number and kind of such branes that occur in (compactified)
string theory.

The first case to consider is (toroidally compactified) IIA/IIB string theory.
Our search for branes will be guided by using as input
supergravity as a low-energy approximation to string theory.
Often, the presence of a $p$-brane in string theory can be deduced
from the presence of a rank $(p+1)$-form potential in the
corresponding supergravity theory. We therefore need to know the possible potentials in
supergravity. We will make here use of the relatively new insight
that these potentials are not only the
ones that describe the physical degrees of freedom of the
supermultiplet. It turns out that the supersymmetry algebra allows
additional high-rank potentials \cite{Riccioni:2007au,Bergshoeff:2007qi,deWit:2008ta}
that do not describe any degree of
freedom but, nevertheless, play an important role in describing the
coupling of branes to the background fields.

Using the potentials of supergravity as input we have started in our earlier work \cite{Bergshoeff:2010xc,arXiv:1102.0934,arXiv:1108.5067}
a classification programme of branes that is based on
considering T-duality covariant and gauge-invariant expressions for the Wess-Zumino (WZ) terms that describe
the coupling of these  branes to the background fields. It is convenient to
classify the branes according to the way their brane tension $T$ scales with the string coupling constant.
This can be expressed by a number $\alpha$ which in string fame is defined by
\begin{equation}\label{def}
T\ \ \sim \ \ (g_s)^{\alpha}\,.
\end{equation}
The branes with $\alpha=0,-1,-2$ are called fundamental, Dirichlet and solitonic branes, respectively.

To determine whether a given potential couples to a half-supersymmetric brane or not
we impose the following
{\sl half-supersymmetric brane criterion}\,:
a potential can be associated to a half-supersymmetric brane
if the corresponding gauge-invariant WZ term requires the
introduction of world-volume fields that fit within the bosonic
sector of a suitable supermultiplet with 16 supercharges.
Imposing this  criterion we have classified all half-su\-per\-sym\-me\-tric branes with $\alpha=0,-1,-2,-3$.
Remarkably, we found that for $\alpha=-2,-3$ not all T-duality representations of potentials
correspond to half-supersymmetric branes and, moreover, that even within a given
T-duality representation not each component corresponds to a half-supersymmetric brane.
We found that this phenomenon only does happen for branes with less than or equal to two
transverse directions. These branes are non-standard in the sense that they are not asymptotically flat
and, furthermore, to obtain finite-energy configurations, one needs to consider multiple branes and to introduce
orientifolds. Here we only focus on single branes that satisfy the half-supersymmetric brane criterion. The non-standard branes
can be divided into defect branes (two transverse directions), domain-walls (one transverse direction) and
space-filling branes (no transverse direction).

We found in our previous work that the potentials that couple to the half-super\-sym\-metric branes
with $\alpha=0,-1,-2,-3$, occur in universal T-duality representations that are valid for any dimensions $3\le D\le 10$,
see Table 1. In each T-duality representation there is always at least one component corresponding to
the reduction of a IIA/IIB brane.
To determine which components of these T-duality representations actually correspond to a
half-supersymmetric brane we must follow a simple  selection rule that will be discussed in the next section.
Remarkably, all these branes can be obtained  from the ten-dimensional branes by a set of wrapping rules which
are also given in Table 1.

\begin{table}\small
\begin{center}
\begin{tabular}{|c||c|c||c|c|}
\hline \rule[-1mm]{0mm}{6mm} $\alpha$ & tensors  & tensor-spinors & wrapped & unwrapped\\
\hline
\hline \rule[-1mm]{0mm}{6mm}
$0$ & $B_{1,A}\,, B_2$& --& doubled & undoubled\\[.1truecm]
\hline \rule[-1mm]{0mm}{6mm}
$-1$ & -- & $C_{2n+1,a}\,, C_{2n,{\dot a}}$& undoubled & undoubled\\[.1truecm]
\hline \rule[-1mm]{0mm}{6mm}
$-2$ & $D_{D-4}\,, D_{D-3,A}\,,$&--&undoubled & doubled\\[.1truecm]
&$D_{D-2,A_1 A_2}\,, D_{D-1,A_1 A_2 A_3}\,, D_{D,A_1 A_2 A_3 A_4}$&--& undoubled & doubled\\[.1truecm]
\hline \rule[-1mm]{0mm}{6mm}
$-3$ & --&$E_{D-2, {\dot a}}\,, E_{D-1, A{\dot a}}\,, E_{D,A_1 A_2 {\dot a}}$& doubled & doubled\\[.1truecm]
\hline \rule[-1mm]{0mm}{6mm}
$-4$ & $F^+_{D,A_1\cdots A_d}$ & --& doubled & --\\[.1truecm]
\hline
\end{tabular}
\end{center}
  \caption{\sl Universal T-duality representations and wrapping rules for all branes in $D$ dimensions that contain amongst the T-duality components at least one brane that follows from the reduction of a brane of IIA/IIB string theory.  Such branes satisfy the wrapping rules given in the last two columns.
  Capital indices $A$ refer to vector indices of the T-duality group $\text{SO}(d,d)$ with $d=10-D$. Repeated vector indices form anti-symmetric tensor representations. The indices $a,\dot a$ refer to chiral and
  anti-chiral spinor indices.
  \label{table1}}
\end{table}

In the first part of this work we will finish the programme started in \cite{Bergshoeff:2010xc,arXiv:1102.0934,arXiv:1108.5067}
and classify the T-duality representations of the branes with $\alpha=-4,-5,-6$. All other branes, with more
negative values of $\alpha$, can be obtained by the ones with $\alpha =0, \cdots ,-6$ by applying S-duality, see the next section. The $\alpha=-4$ branes are special in the sense that there is a ten-dimensional  $\alpha=-4$ brane. It is the S-dual of the D9-brane. We find that all branes
that follow from the reduction of this particular ten-dimensional S-dual D9-brane can be obtained by a ``last wrapping rule'', see Table 1.
At the same time there are also $\alpha=-4$ branes that do not follow from the reduction of any ten-dimensional brane. The same applies to all
$\alpha=-5$ and $\alpha= -6$ branes. All these branes will be discussed in the next section.  Although these branes
do not satisfy any simple wrapping rule, we  find that they do occur in universal T-duality representations, see Table 2. The only
exception is the set of  $\alpha=-6$ branes in $D=3,4$. To determine which components
of these T-duality representations correspond to a half-supersymmetric brane we have to apply the same  selection rule that we will use for the $\alpha=-2,-3$ branes.

\begin{table}
\begin{center}
\begin{tabular}{|c||c||c|c|}
\hline \rule[-1mm]{0mm}{6mm} $\alpha$ & tensors  & tensor-spinors\\
\hline
\hline \rule[-1mm]{0mm}{6mm}
$-4$ & $F_{D-1,A_1\cdots A_{d-3}}\,, F_{D,A,B_1\cdots B_{d-3}}$\,, & --\\[.1truecm]
&$F_{D-2, A_1\cdots A_{d-6}}\,, F_{D-1,A,B_1\cdots B_{d-6}}$&--\\[.1truecm]
\hline \rule[-1mm]{0mm}{6mm}
$-5$ & -- & $G_{D,A_1\cdots A_{d-4},{\dot a}}\,,$ \\[.1truecm]
&--&$G_{D-1,A_1\cdots A_{d-6},a}\,, G_{D,A,B_1\cdots B_{d-6},a}$\\[.1truecm]
\hline \rule[-1mm]{0mm}{6mm}
$-6$ & $D=4$\,: $H_{4,A_1 A_2 A_3 A_4}$&--\\[.1truecm]
&$D=3$\,: $H_{2,A_1 A_2 A_3 }\,,H_{3,A,B_1\cdots B_5}$&--\\[.1truecm]
\hline
\end{tabular}
\end{center}
  \caption{\sl Universal T-duality representations for all branes in $D$ dimensions, suggested by supergravity, that contain amongst their T-duality components  none brane that follows from the reduction of a brane of IIA/IIB string theory. These branes
  do not satisfy any simple wrapping rule. Our index notation is explained in the caption of Table \ref{table1}.
  Tensor representations with both $A$ and $B$ indices, separated by a comma, refer to mixed-symmetry representations, see section 2.
  \label{table2}}
\end{table}

Having finished the classification of the half-supersymmetric
branes, we continue in the second part of this work with a detailed
study of the  orbits of T-duality and U-duality to which the charges
of the half-supersymmetric branes belong. Such orbits turn out
always to contain the highest weight of the relevant T-duality or
U-duality representation; they will be referred to as highest weight
orbits. They are always unique, and no further so-called
stratification occurs. Besides the orbits we will also determine the
constraints satisfied by the charges. It turns out that these
constraints are always maximal, in the sense that the number of
constraints is the  maximal one compatible with a non-trivial
representation. This is a general feature of highest weight orbits.
For U-duality orbits, our results generalise the ones obtained in
\cite{Ferrara:1997ci,FG-1,Lu:1997bg} for the standard,
asymptotically flat, branes.\smallskip

The organisation of this work is as follows. In section 2 we
consider the T-duality covariant expressions for the WZ terms of the
$\alpha =-4,-5,-6$ branes and determine the universal T-duality
representations given in Table 2. In section 3 we use these results
to count the total number of half-supersymmetric branes and find
agreement with \cite{arXiv:1109.2025} where the half-supersymmtric
branes were counted using $\text{E}_{11}$ \cite{West:2001as}. Next,
in section 4 we discuss the T-duality and U-duality orbit
classification of the half-supersymmetric branes. We furthermore
analyse the set of invariant constraints defining such orbits. Our
conclusive remarks and an outlook to further developments are given
in section 5. An Appendix on the derivation of the stabiliser of the
highest weight orbits concludes the paper.

\section{T-duality-covariant Wess-Zumino Terms}

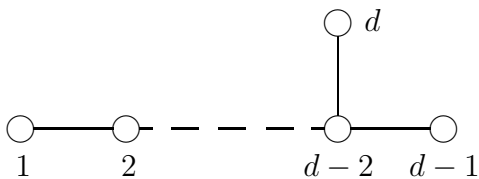
\begin{figure}[h]
\begin{center}
\begin{picture}(180,70)
\multiput(10,10)(40,0){2}{\circle{10}}
\multiput(130,10)(40,0){2}{\circle{10}}
\put(15,10){\line(1,0){30}}
\put(55,10){\line(1,0){5}}
\put(67,10){\line(1,0){10}}
\put(85,10){\line(1,0){10}}
\put(103,10){\line(1,0){10}}
\put(120,10){\line(1,0){5}}
\multiput(135,10)(40,0){1}{\line(1,0){30}} \put(130,50){\circle{10}}
\put(130,15){\line(0,1){30}} \put(8,-8){$1$}
\put(48,-8){$2$} \put(117,-8){$d-2$}
\put(157,-8){$d-1$} \put(140,47){$d$}
\end{picture}
\caption{\sl The $D_d$ Dynkin diagram. \label{DdDynkindiagram}}
\end{center}
\end{figure}

The form fields of maximal supergravity theories in any dimension $D$, including all the so-called {\it non-standard} potentials, that is the forms of rank $D-2$, $D-1$ and $D$, have been classified in \cite{Riccioni:2007au,Bergshoeff:2007qi,deWit:2008ta} according to their U-duality representations. The decomposition of these fields under
 \begin{equation}
  {\rm U-duality} \supset {\mathbb R}^+ \times \text{SO}(d,d)
\quad ,
\end{equation}
where $d=10-D$ and $\text{SO}(d,d)$ is the T-duality symmetry, was listed in \cite{arXiv:1102.0934} for $D\geq 5$ (see Tables 3 to 7 in \cite{arXiv:1102.0934}) and can be easily generalised to dimensions 4 and 3. The $\mathbb{R}^+$-weight of the field is related to the non-positive integer number $\alpha$ defined in eq.~\eqref{def}.
 We already mentioned in the introduction that for non-standard potentials not all the representations, and for a given representation not all its components, are associated to half-supersymmetric branes.  In particular, in \cite{arXiv:1102.0934} and \cite{arXiv:1108.5067} it was shown that, for $\alpha=-2$ and $\alpha=-3$,  only some components of the highest-dimensional representation of a given non-standard form give rise to WZ terms that are compatible with half-maximal supersymmetry. The aim of this section is to generalise and complete this result to all possible values of $\alpha$, determining all the components of T-duality representations of form fields which correspond to half-supersymmetric  branes in any dimension by analysing the field content of their WZ terms.
In order to present the results in the most general way, it is
useful in some cases to give the Dynkin labels of the highest
weights of the T-duality representations we will get with respect to
the nodes of the Dynkin diagram of $\text{D}_d$, see figure
\ref{DdDynkindiagram}. This will be particularly helpful  in section 4,
where the highest weight orbits of these representations will be
determined using the method discussed in the Appendix.

Before we proceed, it is instructive to review the strategy and results of Refs. \cite{arXiv:1102.0934} and \cite{arXiv:1108.5067}.
We started by considering the gauge transformations of the $\alpha =0$ fields (see Table 1)
\begin{equation}
B_{1,A} \quad B_2
\end{equation}
and the $\alpha=-1$ fields (we specify for later convenience here also the Dynkin labels of the highest weight)
\begin{equation}
C_{2n+1, a} \quad (0,0,0,...,0,1)  \quad \qquad C_{2n,\dot{a}} \quad (0,0,0,...,1,0) \quad .
\end{equation}
 Here $A$ is the vector index and $a$ and $\dot{a}$ the two spinor indices of $\text{SO}(d,d)$. Using these transformation rules one can write down the WZ term for the $\alpha =-2$ fields. The supersymmetric branes are then selected with the following criterion: a brane is half-supersymmetric if and only if the worldvolume fields that occur in the leading terms  of the WZ term form the bosonic sector of a half-supersymmetric multiplet. It is important to stress that the worldvolume fields occur in a democratic formulation, that is for instance for a $(p+1)$-dimensional worldvolume a vector field occurs if both a worldvolume vector and a dual  worldvolume $(p-2)$-form appear in the WZ term.\footnote{In the case of a six-dimensional world-volume, when only a single worldvolume 2-form occurs, that form is considered to be self-dual.} The result is that for $\alpha=-2$, all the branes are associated to the form fields (see Table 1)
\begin{equation}\label{susyforms}
D_{D-4} \qquad D_{D-3,A}\qquad D_{D-2,A_1 A_2} \qquad D_{D-1,A_1 A_2 A_3} \qquad D_{D,A_1 A_2 A_3 A_4} \quad ,
\end{equation}
where antisymmetrisation in the vector indices is understood.
Moreover, taking each index along lightlike directions, denoted by $m\pm$ in \cite{arXiv:1102.0934}, we find the restriction that in the case of multiple indices $m\pm n \pm p\pm ...$ the  indices  $m$, $n$, $p$, ... have to be all different.
This gives the following number of non-standard supersymmetric $\alpha=-2$  branes:\,\footnote{For instance, in the case of defect branes, which couple
to the $(D-2)$-forms in eq.~\eqref{susyforms}, the counting is explained as follows: ignoring the $\pm$-indices, a 2-form in the $m$-indices leads to
${d \choose 2}$ components since $m$ takes $d$ values. Next, there are $2^2$ ways to divide the remaining $\pm$ indices over the
two places.}
\begin{equation}
{\rm defect \  branes:}\  {d \choose 2}\times 2^2 \quad {\rm domain \ walls:}\  {d \choose 3} \times 2^3 \quad {\rm space-filling\ branes:} \  {d \choose 4} \times 2^4
\  .
\end{equation}

Similarly, in \cite{arXiv:1108.5067} the same analysis was performed for $\alpha =-3$ branes, showing that they are associated to the form fields
\begin{equation}
E_{D-2,\dot{a}} \ (0,0,0,...,1,0) \qquad E_{D-1, A\dot{a}} \ (1,0,0,...,1,0) \qquad E_{D, A_1 A_2\dot{a}} \ (0,1,0,...,1,0)\,,
\end{equation}
 which are in irreducible spinor, vector-spinor and tensor-spinor representations respectively. For the vector indices the same rule as for the $\alpha=-2$ fields applies, that is the indices are along the lightcone directions and for the case of space-filling branes the two indices $A_1 A_2$ are $m\pm n \pm$ with $m \neq n$. We now supply this with the additional rule that, for each vector index, only half of the spinor indices contributes. This gives the following number of $\alpha=-3$ branes:
\begin{equation}
{\rm defect \  branes:}\  2^{d-1} \quad {\rm domain \ walls:}\  d \times 2 \times 2^{d-2} \quad {\rm space-filling\ branes:} \  {d \choose 2} \times 2^2 \times 2^{d-3}
\  .
\end{equation}

We now want to perform the same analysis for the lower values of $\alpha$, i.e. $\alpha \leq -4$. The lowest value of $\alpha$ is $-11$ and it occurs for space-filling branes in three dimensions. In three dimensions, though, the representations with a given value of $\alpha$ are mapped to the ones with $-\alpha -4 (p+1 )$ by conjugation. This means that we only have to consider $\alpha=-6$ as the lowest value. Similarly, in four dimensions the lowest value of $\alpha $ is $-7$, which occurs for space-filling branes, but again the representations are mapped by S-duality according to $\alpha \rightarrow -\alpha -2 (p+1)$. Finally, in $D= 5$ and $D=6$  the lowest value of $\alpha$ is  $-5$, while for $D \geq 7$ the lowest value of $\alpha$ is always $-4$. Taking everything together, this means that we only have to consider the cases $\alpha=-4$, $\alpha =-5$ and  $\alpha =-6$.   The notation that we use for the target-space fields, the corresponding worldvolume fields and their worldvolume field-strength is summarised in Table \ref{notation}.
\begin{table}
\begin{center}
\begin{tabular}{|c||c|c|c|}
\hline \rule[-1mm]{0mm}{6mm} value of $\alpha$ & t.s. field  & w.v.  field & w.v. field-strength\\
\hline
\hline \rule[-1mm]{0mm}{6mm}
$\alpha=0$ & $B$  & $b$ & ${\cal F}$\\
\hline \rule[-1mm]{0mm}{6mm}
$\alpha=-1$ & $C$  & $c$ & ${\cal G}$\\
\hline \rule[-1mm]{0mm}{6mm}
$\alpha=-2$ & $D$  & $d$ & ${\cal H}$\\
\hline \rule[-1mm]{0mm}{6mm}
$\alpha=-3$ & $E$  & $e$ & ${\cal I}$\\
\hline \rule[-1mm]{0mm}{6mm}
$\alpha=-4$ & $F$  & $f$ & ${\cal J}$\\
\hline \rule[-1mm]{0mm}{6mm}
$\alpha=-5$ & $G$  & $g$ & ${\cal K}$\\
\hline \rule[-1mm]{0mm}{6mm}
$\alpha=-6$ & $H$  & $h$ & ${\cal L}$\\
\hline
\end{tabular}
\end{center}
  \caption{\sl Summary of our notation for the various target-space (t.s.) potentials and their corresponding world-volume (w.v.) potentials and field-strengths.\label{notation}}
\end{table}
Our strategy is now as follows. For each field, we write down all possible leading WZ terms compatible with T-duality and $\alpha$-conservation. We will not compute the actual coefficient of each term, but instead we will assume that all terms that can in principle occur will actually occur with non-zero coefficient. By looking at the structure of the WZ terms, we will select the supersymmetric branes as the ones propagating the correct worldvolume degrees of freedom. We will identify all the fields that are associated to $\alpha=-4$, $\alpha=-5$ and $\alpha=-6$ branes, and we will show that the components of their T-duality representations are determined according to a  selection rule that generalises the rule for $\alpha=-2$ and $\alpha=-3$ as follows:
\bigskip

\noindent {\sl Supersymmetric selection rule\,:}
\vskip .1truecm

\begin{itemize}
\item the T-duality vector indices are all in lightcone directions. The antisymmetric indices are of the form $m\pm n\pm p\pm ...$ with $m$, $n$, $p$,... all different. For mixed-symmetry representations,  each index that is not antisymmetrised has to be parallel to one of the antisymmetric indices;

\item the number of spinor components are halved each time a different vector index appears.
\end{itemize}
In \cite{arXiv:1109.1725} the WZ terms for all the half-supersymmetric branes in  dimensions  $6 \le D \le 9$  were determined using a different method, following the approach of \cite{Bergshoeff:2010xc}. Our results in this paper coincide with those obtained in that paper where they overlap.

In \cite{arXiv:1109.2025} all half-supersymmetric branes  in $3\le D\le 10$ and for any value of $\alpha$ were counted using an algebraic method based on $\text{E}_{11}$ \cite{West:2001as}, which consists in first decomposing $\text{E}_{11}$ in $\text{GL}(D,\mathbb{R}) \times E_{11-D(11-D)}$ and then counting all $\text{E}_{11}$ roots associated to generators that give rise to antisymmetric representations of $\text{GL}(D,\mathbb{R})$, corresponding to forms in $D$ dimensions, and selecting only those that have positive squared length (real roots).
The overall counting of the branes obtained using our WZ term method exactly reproduces the results of \cite{arXiv:1109.2025} in all cases.

Another remarkable feature of our analysis is that, in five dimensions and below, there are some values of $\alpha$ for which two T-duality representations instead of one contain supersymmetric branes of a given worldvolume dimension
and containing the same worldvolume multiplet. In particular, we find that this occurs for the $\alpha=-4$ space-filling branes in five, four and three dimensions, as well as for the $\alpha=-4$ domain walls and the $\alpha=-5$ space-filling branes in three dimensions. We also find that a subsector of the $\alpha=-4$ branes can be obtained starting from the $\alpha=-4$  S-dual of the IIB D9-brane by applying a specific wrapping rule,
see Table \ref{table1}. This generalises the results in \cite{arXiv:1106.0212,arXiv:1108.5067} which showed that all the branes with $\alpha$ from 0 to $-3$ result from the 10-dimensional ones using specific wrapping rules. Since there are no branes in ten dimensions with $\alpha<-4$, there are no further wrapping rules.

We will now proceed with a case by case analysis of the WZ terms of the $\alpha=-4$, $\alpha=-5$ and $\alpha=-6$ branes separately. These results will be summarised and collected for the different dimensions in the next section.  The reader who is not interested in the explicit derivation of the WZ terms below can jump directly to that section.

\subsection{The $\alpha=-4$ Branes and the Last Wrapping Rule}

We start our analysis by considering the branes with $\alpha=-4$, which is the lowest value of $\alpha$ that occurs in ten dimensions, corresponding to  the S-dual of the IIB D9-brane.
The WZ term of this brane has the form
\begin{equation}
F_{10} + E_8 {\cal G}_2 + C_2 {\cal I}_8
\end{equation}
corresponding to a vector multiplet in 10 dimensions because the WZ term contains a worldvolume vector $c_1$ and its magnetic dual 7-form $e_7$.

It turns out that what generalises this field in all dimensions is the field $F_{D, A_1 ...A_d}^+$  in the self-dual representation of T-duality with $d$ antisymmetric indices denoted by the Dynkin labels $(0,0,0,...,2,0)$.  This $\alpha=-4$ field is present in all maximal-supergravity theories: it is the 8-form in the  ${\bf (3,1)}$  in eight dimensions, the 7-form in the  ${\bf 10}$ in seven dimensions and so on.
The general expression for the WZ term for this field is
\begin{eqnarray}
& & F_{D, A_1 ...A_{d}}^+ + E_{D-2, \dot{a}} {\cal G}_{2 ,\dot{b}} (C \Gamma_{A_1 ...A_d } )^{\dot{a}\dot{b}}  + E_{D-1 , [A_1 , \dot{a}} {\cal G}_{1, a} (C \Gamma_{A_2 ...A_d ]} )^{a \dot{a}} \nonumber \\
& & + C_{2, \dot{a}} { \cal I}_{D-2, \dot{b}} (C \Gamma_{A_1 ...A_d } )^{\dot{a}\dot{b}}+ C_{1,a} {\cal I}_{D-1  [A_1 , \dot{a}} (C \Gamma_{A_2 ...A_d ]} )^{a \dot{a}} \quad ,
\end{eqnarray}
where self-duality on the $A_1 ...A_d$ indices is understood in all terms.
This expression contains  in all dimensions one vector (the Gamma matrix with $d$ indices selects one component out of a $2^{d-1}$-dimensional spinor) and $d$ scalars (one for each $A$ index), which makes the bosonic sector of a vector multiplet in $D$ dimensions. Using our supersymmetric selection rule, we can count the number of such branes. The result is
  \begin{equation}
 \frac{1}{2} \times {d \choose d } \times 2^d = 2^{d-1} \quad ,
\end{equation}
where the first factor $1/2$ follows from self-duality.
Remarkably, all these branes can be obtained from 10 dimensions by means of the wrapping rule \,\footnote{Note that a space-filling brane can only wrap.}
  \begin{equation}\label{newwrapping}
  {\rm wrapped} \ \ \ \ \rightarrow\  \ \ {\rm doubled}\, ,
 \end{equation}
where when going from ten to nine the doubling means that one considers both the branes coming from IIA and from IIB (and there is no such brane in the IIA theory). The result is summarised in Table \ref{alpha-4table}.
\begin{table}[h]
\begin{center}
\begin{tabular}{|c||c|c|c|c|c|c|c|c|}
\hline \rule[-1mm]{0mm}{6mm} $p$-brane &IIA/IIB& 9 & 8 & 7 & 6&5&4&3\\
\hline
 \hline \rule[-1mm]{0mm}{6mm} 2&&&&&&&&64\\
 \hline \rule[-1mm]{0mm}{6mm} 3&&&&&&&32&\\
 \hline \rule[-1mm]{0mm}{6mm} 4&&&&&&16&&\\
 \hline \rule[-1mm]{0mm}{6mm} 5& &&&&8&&&\\
 \hline \rule[-1mm]{0mm}{6mm} 6& &  &  &4&&&&\\
 \hline \rule[-1mm]{0mm}{6mm} 7& &  & 2 & & &&&\\
 \hline \rule[-1mm]{0mm}{6mm} 8& & 1 &  & & &&&\\
 \hline \rule[-1mm]{0mm}{6mm} 9& 0/1&  &  & & &&&\\
\hline
\end{tabular}
\caption{\sl By applying the wrapping rule (\ref{newwrapping})  one
obtains a subset of the $\alpha =-4$ supersymmetric space-filling branes, which are the ones associated to the fields $F_{D, A_1 ...A_d}^+$.
\label{alpha-4table}}
\end{center}
\end{table}
This result shows that all supersymmetric branes of IIA and IIB ten-dimensional string theory satisfy wrapping rules upon dimensional reduction.

A closer look at the fields with $\alpha=-4$, as well as all the other fields in Table \ref{notation} that can occur in the WZ term, reveals that there is a new set of fields giving rise to domain walls from seven dimensions downwards. This domain wall, whose dynamics is described
by a worldvolume tensor multiplet, does not follow from a wrapped brane in ten dimensions.
The technical reason for the occurrence of this domain wall is that in seven dimensions one can introduce the term $D_3 {\cal H}_3$ which is a singlet in a domain wall WZ term. This term corresponds to a self-dual tensor. More specifically, there is a singlet $\alpha=-4$ 6-form field in seven dimensions such that its WZ term is
\begin{equation}
F_6 + E_{5 ,\dot{a}}{ \cal G}_{1 , a} C^{a \dot{a}} + C_{1 ,a } {\cal I}_{5 \dot{a}} C^{a \dot{a}} + D_3 {\cal H}_3  \quad .
\end{equation}
This describes four scalars, one self-dual tensor together with one transverse scalar, which corresponds to a tensor multiplet in six dimensions. This WZ term was already considered in \cite{arXiv:1109.1725} as can be seen in the second line of Table 6 of that paper.
This seven-dimensional domain wall generalises to other domain walls in any dimension below seven. They are collectively described by the field $F_{D-1, A_1 ...A_{d-3}}$ which is for instance in  the ${\bf 45}$ in five dimensions. Its WZ term is
 \begin{eqnarray}
& & F_{D-1, A_1 ... A_{d-3} } + E_{D-2, \dot{a}} {\cal G}_{1,a} (C \Gamma_{A_1 ...A_{d-3}} )^{a \dot{a}} + D_{D-4} {\cal H}_{3, A_1 ...A_{d-3}} \nonumber \\
& & + D_{D-3, [A_1} {\cal H}_{2, A_2 ... A_{d-3}]} + D_{D-2, [A_1 A_2} {\cal H}_{1 , A_3 ...A_{d-3}]} \quad .
\end{eqnarray}
Picking the lightcone directions according to our selection rule, one can show that in all cases this contains a vector multiplet.  We will perform the detailed counting of the number of corresponding branes in the next section.

The $(D-1)$-form potentials just considered induce a new set of space-filling branes, in dimension less than or equal to six, in the mixed-symmetry representation $F_{D, A , B_1 ... B_{d-3}}$ where the $B$ indices are antisymmetrised. The highest dimension in which such a brane appears is $D=6$. The corresponding WZ term is given by
  \begin{eqnarray}
& & F_{6, AB} + E_{5 , (A, \dot{a}} {\cal G}_{1, a} (C \Gamma_{B)})^{a \dot{a}} +  D_{3 ,(A} {\cal H}_{3 , B)} + F_{5 , (A} {\cal F}_{1, B)} \nonumber \\
& & + {C}_{1, a}  {\cal I}_{5 , (A, \dot{a}}  (C \Gamma_{B)})^{a \dot{a}} + B_{1, (A}  {\cal J}_{5 , B)} \quad ,
\end{eqnarray}
with $AB$ symmetric. Using our selection rule, the indices $A$ and $B$ have to be parallel and along a lightlike direction. One can see that this implies a single self-dual tensor and five scalars, corresponding to a six-dimensional tensor multiplet.  Observe that one could have  in principle added  the term $E_{4 ,\dot{a}} {\cal G}_{2, \dot{b}} C^{\dot{a} \dot{b}} \eta_{AB} $, which would have ruined the counting. This term, however,  vanishes because $\eta_{AB}$ is off-diagonal in lightcone coordinates. This tensor brane was already determined in \cite{arXiv:1109.1725} (first line of Table 6 of that paper).
It can be shown that this six--dimensional space-filling brane generalises to all dimensions below six, with worldvolume fields describing a vector multiplet. The counting of these branes will be performed in the next section.

Starting from four dimensions a new set of defect branes occurs, described by the field  $F_{D-2, A_1 ... A_{d-6}}$. Actually, since we only discuss dimensions higher than 2, this will only be relevant in dimensions 4 and 3. For instance, the WZ term of the four-dimensional defect brane is
  \begin{equation}
F_{2} + D_{1, A} {\cal H}_1^A  \quad ,
\end{equation}
describing six scalars (that is 12 scalars satisfying self-duality conditions) plus two transverse scalars. This makes a total of eight scalars, which form the bosonic sector of a two-dimensional scalar multiplet.

The above set of $D=4$  defect branes induces a new set of $D=3$ domain walls described by the  fields $F_{D-1, A, B_1 ...B_{d-6}}$. In three dimensions this is a symmetric tensor  giving  the WZ term
  \begin{equation}
F_{2, AB} + F_{1, (A} {\cal F}_{1, B)} + D_{1, (C(A} {\cal H}_{1, B)D} \eta^{CD} + B_{1, (A} { \cal J}_{1,B)}
\end{equation}
with $AB$ symmetric.
Taking for instance the direction $m \hskip -.1truecm + m+$, this contains two dual scalars $b_{0, m+}$ and $f_{0, m+}$, and  12 scalars $d_{0,m+ n\pm}$, with $n \neq m$ taking only 6 values for fixed $m$. These 12 scalars satisfy a self-duality condition. Together with the transverse scalar, this makes a total of eight scalars forming a two-dimensional scalar multiplet.

%One can show that it is not possible to obtain supersymmetric WZ terms for the other fields with $\alpha=-4$ in all cases.

\subsection{The $\alpha =-5$ Branes}
We next proceed determining all possible $\alpha=-5$ branes. We do the analysis in full generality, although we already know the result
in six dimensions and in four dimensions  and for the 1-branes in three dimensions because of S-duality.

The highest dimension in which space-filling branes with $\alpha =-5 $  occur is $D=6$. The corresponding field is $G_{6, \dot{a}}$, giving 8 branes in the spinor representation of $\text{SO}(4,4)$ denoted by the Dynkin labels $(0,0,1,0)$. The WZ term is
  \begin{equation}
  G_{6 ,\dot{a}} + E_{4, \dot{a}} {\cal H}_2 +D_2 {\cal I}_{4, \dot{a}}+  F_{5, A}  {\cal G}_{1, a }  \Gamma^A_{\dot{a}}{}^a + C_{1, a } {\cal J}_{5,A} \Gamma^A_{\dot{a}}{}^a\,,
\end{equation}
which describes, for a fixed value of $\dot a$, a vector $d_1$ and its dual $e_3$, together with four scalars $c_0$ and their dual $f_4$, which makes a vector multiplet in six dimensions.
In lower dimensions, this space-filling brane generalises to a whole family of space-filling branes described by the field $  G_{D, A_1 ...A_{d-4} \dot{a}}$, with the $A$ indices antisymmetrised. The corresponding  WZ term is given by
  \begin{eqnarray}
& &    G_{D, A_1 ...A_{d-4} \dot{a}} +  E_{D-2, \dot{a}} {\cal H}_{2, A_1 ... A_{d-4}} + E_{D-1, [A_1} {\cal H}_{1, A_2 ...A_{d-4}]} + F_{D-1, [A_1 ...A_{d-5}} {\cal G}_{1,a}\Gamma_{A_{d-4}],\dot{a}}{}^a  \nonumber \\
& & + D_{2, A_1 ... A_{d-4}} {\cal I}_{D-2, \dot{a}} + D_{1 , [ A_1 ...A_{d-5}} {\cal I}_{D-1, A_{d-4}]} + C_{1,a} {\cal J}_{D-1  [A_1 ...A_{d-5}} \Gamma_{A_{d-4}],\dot{a}}{}^a
 \end{eqnarray}
 describing a vector multiplet in $D$ dimensions.

Starting from four dimensions, there is a new chain of $(D-1)$-form fields giving rise to supersymmetric WZ terms for domain walls. The fields are $G_{D-1, A_1 ...A_{d-6},a }$. The WZ term in four dimensions is given by
  \begin{equation}
G_{3, a } +  F_2 {\cal G}_{1,a} + E_{2,\dot{a}} {\cal H}_{1, A} \Gamma_{A, a }{}^{\dot{a}}  + C_{1,a} {\cal J}_2 + D_{1,A} {\cal I}_{2,\dot{a}}  \Gamma_{A, a }{}^{\dot{a}}
\end{equation}
describing a scalar multiplet in three dimensions. Similarly, one can show that the field $G_{2,Aa}$ in three dimensions describes a  supersymmetric domain wall.
This last brane induces an additional space-filling brane with $\alpha=-5$ in three dimensions, associated to the field $G_{3,ABa}$ symmetric in $AB$.
 The WZ term is
 \begin{eqnarray}
& & G_{3, ABa} +  F_{2,AB} {\cal G}_{1, a} + G_{2, (A,a} {\cal F}_{1, B)} + E_{2, (A, \dot{a}} {\cal H}_{1, B)C} \Gamma^C_a{}^{\dot{a}} + C_{1,a} {\cal J}_{2,AB}\nonumber \\
& &  + B_{1,(A} {\cal K}_{2, B)a} + D_{1,C(A} {\cal I}_{2, B)\dot{a}} \Gamma^C_a{}^{\dot{a}}
\end{eqnarray}
describing a scalar multiplet.
This concludes the analysis of the $\alpha=-5$ branes.

\subsection{The $\alpha =-6$ Branes}
We finally consider the $\alpha=-6$ branes. The $\alpha=-6$  space-filling brane in four dimensions is the S-dual of the $\alpha =-2$ one, and  the $\alpha=-6 $ domain wall in three dimensions is similarly the S-dual of the $\alpha=-2$ one. We thus only need to consider the $\alpha=-6$ space-filling branes in three dimensions.
The only three-dimensional field with $\alpha =-6$ that gives rise to the right number of worldvolume degrees of freedom is  $H_{3, A, B_1 ...B_5}$ in the mixed-symmetry representation with the $ B$ indices antisymmetrised.  The WZ term is
\begin{eqnarray}
& & H_{3, A, B_1 ...B_5} + F_{2, [B_1 ....B_4} {\cal H}_{1, B_5 ]A} + E_{2, A\dot{a}} {\cal I}_{1,\dot{b}} (C \Gamma_{B_1 ... B_5} )^{\dot{a}\dot{b}} + G_{2, Aa} {\cal G}_{1, b} (C \Gamma_{B_1 ...B_5})^{ab} \nonumber \\
 & & + D_{1, A [B_1} {\cal  J}_{2, B_2....B_5 ] } + {E}_{1,\dot{a}}  {\cal I}_{2, A\dot{b}}  (C \Gamma_{B_1 ... B_5} )^{\dot{a}\dot{b}} + C_{1,a} {\cal K}_{2, Ab} (C \Gamma_{B_1 ...B_5})^{ab}\,,
\end{eqnarray}
which describes a scalar multiplet. This finishes our discussion of the $\alpha=-6$ branes.

\vskip .5cm

One could in principle perform the same analysis for all the other values of $\alpha$, but this is not necessary because they are all related to known cases by S-duality. In the next section we will proceed with the counting of the supersymmetric branes.

\section{Brane Counting}
In this section we will perform the counting of the non-standard branes that result from the analysis of the previous section. Given that the WZ terms of supersymmetric branes in all dimensions above five were already determined in \cite{arXiv:1109.1725}, we will only perform this analysis in five, four and three dimensions.
As already stressed, the overall number of supersymmetric branes that we obtain for each value of $\alpha$ has already been determined in \cite{arXiv:1109.2025} using a different approach based on $\text{E}_{11}$ \cite{West:2001as}. More precisely, the method used in \cite{arXiv:1109.2025} is based on counting all the form fields in a given dimension that correspond to $\text{E}_{11}$ roots with positive squared length (that is real roots). Here we will show that if one considers the forms in a given representation of T-duality such that its highest weight corresponds to a real root of $\text{E}_{11}$, then it is enough to count for that representation all the components that satisfy our supersymmetric selection rule. Indeed, all the representations that we selected in the previous section by imposing that the WZ term contains a supersymmetric multiplet are precisely all the representations such that its highest weight is associated to an $\text{E}_{11}$ real root.\,\footnote{This has been checked in all cases using the  simpLie software of \cite{Bergshoeff:2007qi}.}

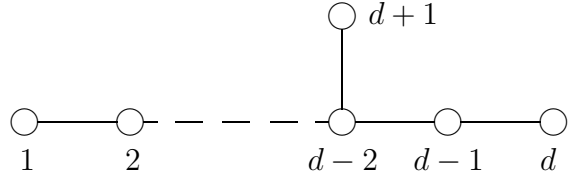
\begin{figure}[h]
\begin{center}
\begin{picture}(220,70)
\multiput(10,10)(40,0){2}{\circle{10}}
\multiput(130,10)(40,0){3}{\circle{10}}
\put(15,10){\line(1,0){30}}
\put(55,10){\line(1,0){5}}
\put(67,10){\line(1,0){10}}
\put(85,10){\line(1,0){10}}
\put(103,10){\line(1,0){10}}
\put(120,10){\line(1,0){5}}
\multiput(135,10)(40,0){2}{\line(1,0){30}} \put(130,50){\circle{10}}
\put(130,15){\line(0,1){30}} \put(8,-8){$1$}
\put(48,-8){$2$} \put(117,-8){$d-2$}
\put(157,-8){$d-1$} \put(205,-8){$d$} \put(140,47){$d+1$}
\end{picture}
\caption{\sl The $E_{d+1(d+1)}$ Dynkin diagram. \label{Ed+1diagram}}
\end{center}
\end{figure}

As far as U-duality is concerned, for $D\leq 5$ all these T-duality representations belong to the highest-dimensional U-duality representation for a given form, and for each form  this U-duality representation is the only one such that its highest weight is associated to a real root. For $D \geq 6$, there is always the special case of the 6-form, for which there are in all cases two different U-duality representations whose highest weight is associated to an $\text{E}_{11}$ real root. It turns out that there is a universal way of classifying the U-duality representations of the form fields such that their highest weight is associated to an  $\text{E}_{11}$ real root in all dimensions. Labelling the nodes of the Dynkin diagram of the $\text{E}_{d+1(d+1)}$ U-duality group as in fig. \ref{Ed+1diagram},
the representations of such  form fields are, in terms of Dynkin indices,\footnote{The fact that for low rank the U-duality representations follow a regular pattern was already observed in \cite{Riccioni:2008jz} (see the Appendix of that paper). Here we emphasise that this regularity is universal if one only concentrates on the fields whose highest weights correspond to real roots of $\text{E}_{11}$.}
  \begin{eqnarray}
& & {\rm 1-form}: \quad (1,0,...,0,0,0,0) \nonumber \\
& & {\rm 2-form}: \quad (0,0,...,0,0,1,0)\nonumber \\
& &  {\rm 3-form}: \quad (0,0,...,0,0,0,1)\nonumber \\
& &   {\rm 4-form}: \quad (0,0,...,0,1,0,0)\nonumber \\
&  & {\rm 5-form}: \quad (0,0,...,0,0,1,1)\nonumber \\
& &  {\rm 6-form}: \quad (0,0,...,0,1,1,0) \oplus (0,0,...,0,0,0,2)\nonumber \\
&&  {\rm 7-form}: \quad (0,0,...,0,0,2,1) \quad . \label{Udualityreprrealroot}
\end{eqnarray}
The reader should keep in mind that this general notation, applied to the seven-dimensional case, that is $d=3$, and the six-dimensional case, that is $d=4$, gives  a labelling of the nodes of the $\text{A}_4$ and $\text{D}_5$ Dynkin diagrams which is not the standard one.
We limit here the analysis to rank 7 and lower because the 8-forms only occur in eight, nine and ten dimensions, and this makes our general analysis less meaningful, due to the pathology of the Dynkin diagram of fig. \ref{Ed+1diagram} for  $d=1$ and $d=0$, correspondiing to nine and ten dimensions.
 We will make use of this way of labelling the U-duality representations in the next section, where using the method explained in the Appendix we will determine all their highest weight orbits.
The reader can appreciate that one recovers immediately from eq. \eqref{Udualityreprrealroot} the representations of all the forms whose highest weight is associated to a real $\text{E}_{11}$ root in all dimensions below nine.
The only exception are the 8-forms in eight dimensions, which belong to the ${\bf (15,1)}$ of $\text{SL}(3,\mathbb{R}) \times \text{SL}(2,\mathbb{R})$, which is $(1,2)\oplus 0$ in terms of the Dynkin diagram of fig. \ref{Ed+1diagram} for $d=2$. Also, again in $D=8$ the case of the 1-forms is exceptional, because they belong to the ${\bf (\overline{3},2)}$, corresponding to Dynkin labels $(1,0) \oplus 1$, and therefore stricktly speaking the rule of eq. \eqref{Udualityreprrealroot} does not apply. This can be easily understood from $\text{E}_{11}$. Taking the $\text{E}_{11}$ Dynkin diagram (which is just the diagram in fig.  \ref{Ed+1diagram} for $d=10$) one obtains the $D$-dimensional theory by decomposing $\text{E}_{11}$ in $\text{GL}(d,\mathbb{R} ) \times \text{E}_{d+1(d+1)}$, which corresponds to deleting node $D$. The representations of the 1-forms can then be read by looking at the nodes of $\text{E}_{d+1(d+1)}$ that are connected to the deleted node \cite{Kleinschmidt:2003mf}. This gives the 1-forms with Dynkin labels  given in eq. \eqref{Udualityreprrealroot} for $D \leq 7$, and $(1,0) \oplus 1$ for $D=8$.

We now perform the analysis of the non-standard branes with $\alpha =-4$, $-5$ and $-6$  in detail starting from $D=5$.

\begin{table}
\begin{center}
\begin{tabular}{|c|c||c|c|c|c|c|}
\hline \rule[-1mm]{0mm}{6mm} brane & U repr  & $\alpha=-1$ &
$\alpha=-2$ & $\alpha=-3$ & $\alpha=-4$ & $\alpha=-5$\\
\hline
\hline \rule[-1mm]{0mm}{6mm}
2-brane & $72\subset {\bf {78}}$  & $16 \subset{\bf 16}$ & $40\subset {\bf 45} $ & $16 \subset {\bf \overline{16}} $&&\\
\hline \rule[-1mm]{0mm}{6mm}
3-brane & $216\subset {\bf {351}}$  & $ 16 \subset {\bf \overline{16}}$ & $80 \subset {\bf 120}$ &
$ 80\subset {\bf 144}
$ & $40 \subset {\bf 45}$ &\\
\hline \rule[-1mm]{0mm}{6mm}
4-brane & $432 \subset {\bf \overline{1728}}$  &  $16 \subset {\bf
{16}}$ & $80 \subset {\bf 210}$ & $160 \subset {\bf \overline{560}}$ & $80 \subset  {\bf 320}  $ & $80 \subset {\bf 144}$  \\
& & & & &  $16 \subset {\bf \overline{126}}$ &  \\
\hline
\end{tabular}
\end{center}
  \caption{\sl The  non-standard branes of $D=5$ maximal
supergravity. The U-duality group is $\text{E}_{6(6)}$ and
the T-duality group is $\text{SO}(5,5)$.\label{D=5branescan}}
\end{table}

\subsection{$D=5$}
In five dimensions the U-duality symmetry is $\text{E}_{6(6)}$, and the T-duality symmetry is $\text{SO}(5,5)$.
The only branes with $\alpha <-3$ are the domain walls and the space-filling branes. For the domain walls we get from the previous section
the $\alpha=-4$ field
$  F_{4, A_1 A_2}$ in the   ${\bf 45}$ of $\text{SO}(5,5)$,
giving rise to
\begin{equation}
 F_{4 , m_1 \pm m_2 \pm} \ \   \rightarrow {5 \choose 2} \times 2^2 = 40  \quad .
\end{equation}
For the space-filling branes we get the $\alpha=-4$ fields $F_{5, A_1 ...A_5}^+$ in the ${\bf \overline{126}}$ and $F_{5, A, B_1 B_2}$ in the ${\bf 320}$, giving
  \begin{eqnarray}
 & &  F_{5, m_1 \pm ...m_5\pm}^+ \rightarrow \frac{1}{2}\times  {5\choose 5}\times 2^5 = 16 \quad , \nonumber \\
& & F_{5, m_1 \pm, m_1 \pm m_2\pm} \rightarrow {5 \choose 2} \times 2^2 \times 2 = 80 \quad ,
\end{eqnarray}
together with the $\alpha=-5$ field $G_{5, A\dot{a}}$ in the ${\bf 144}$, giving
  \begin{equation}
 G_{5 , m\pm \dot{a}} \rightarrow 5 \times 2 \times 2^{5-2} = 80 \quad .
  \end{equation}
The result (including the cases with higher $\alpha$) is summarised in Table \ref{D=5branescan}.
%We stress again that for each of these forms, the highest weight of the representations we have considered are the %only ones that are associated with $\text{E}_{11}$ roots with positive squared length.
 All the space-filling branes belong to the ${\bf \overline{1728}}$ of $\text{E}_{6(6)}$. This is the first example in which a given U-duality representation gives rise to two different sets of branes, with the same worldvolume multiplet, for a given $\alpha$ (namely $\alpha=-4$) corresponding to different T-duality representations.

\subsection{$D=4$}

\begin{table}\scriptsize
\begin{center}
\begin{tabular}{|c|c||c|c|c|c|c|c|c|c|}
\hline \rule[-1mm]{0mm}{6mm} brane & U repr & $\alpha=0$   & $\alpha=-1$ &
$\alpha=-2$ & $\alpha=-3$ & $\alpha=-4$ & $\alpha=-5$ & $\alpha=-6$ & $\alpha =-7$\\
\hline
\hline \rule[-1mm]{0mm}{6mm}
1-brane & $126\subset {\bf {133}}$  & $1 \subset{\bf 1}$  & $32 \subset {\bf \overline{32}}$
& $66\subset {\bf 66} $ & $32 \subset {\bf \overline{32}} $& $1 \subset {\bf 1}$ & & &  \\
\hline \rule[-1mm]{0mm}{6mm}
2-brane & $576\subset {\bf {912}}$  &  & $ 32 \subset {\bf {32}}$ & $160 \subset {\bf 220}$ &
$ 192\subset {\bf 352}
$ & $160 \subset {\bf 220}$ & $32 \subset {\bf 32}$ & & \\
\hline \rule[-1mm]{0mm}{6mm}
3-brane & $2016 \subset {\bf {8645}}$  &  & $32 \subset {\bf
\overline{32}}$ & $240 \subset {\bf 495}$ & $480 \subset {\bf \overline{1728}}$ & $480 \subset  {\bf 2079}$ & $480 \subset {\bf \overline{1728}}$ & $240 \subset {\bf 495}$  & $32 \subset {\bf \overline{32}}$   \\
 & & & & & &  $32 \subset {\bf \overline{462}}$  & & & \\
\hline
\end{tabular}
\end{center}
  \caption{\sl The  non-standard branes of $D=4$ maximal
supergravity. The U-duality group is $\text{E}_{7(7)}$ and
the T-duality group is $\text{SO}(6,6)$.\label{D=4branescan}}
\end{table}

In four dimensions the U-duality symmetry is $\text{E}_{7(7)}$ and the T-duality symmetry is $\text{SO}(6,6)$. From the previous section we know that there is a singlet $\alpha=-4$ 2-form $F_2$ giving rise to a defect brane,
while for the domain walls one has the $\alpha=-4$ field $F_{3 , A_1 A_2 A_3}$ in the ${\bf 220}$ and the $\alpha =-5$ field $G_{3, a}$ in the ${\bf 32}$, giving
  \begin{eqnarray}
& & F_{3 , m_1\pm m_2\pm m_3\pm}\rightarrow {6\choose 3} \times 2^3 = 160 \quad , \nonumber \\
& & G_{3,a } \rightarrow 32 \quad .
\end{eqnarray}
Finally, the fields corresponding to space-filling branes are the $\alpha=-4$ fields $F_{4 , A_1 ...A_6}^+$ in the ${\bf \overline{462}}$ and $F_{4, A, B_1 B_2 B_3}$ in the ${\bf 2079}$ giving
  \begin{eqnarray}
& &  F_{4, m_1 \pm m_2 \pm ... m_6 \pm}^+ \rightarrow \frac{1}{2} \times {6 \choose 6} \times 2^6 = 32 \quad ,\nonumber \\
& & F_{4, m_1\pm, m_1\pm m_2\pm m_3\pm } \rightarrow {6\choose 3} \times 2^3 \times 3= 480 \quad .
\end{eqnarray}
All the other values of $\alpha$ can be determined using S-duality. All the space-filling branes belong to the ${\bf 8645}$ of $\text{E}_{7(7)}$. The overall result is summarised in Table \ref{D=4branescan}.
As in the five-dimensional case, the $\alpha=-4$ space-filling branes belong to two different representations of T-duality.

\subsection{$D=3$}

In three dimensions the U-duality is $\text{E}_{8(8)}$, and the T-duality is $\text{SO}(7,7)$. From the analysis of the previous section we have the $\alpha=-4$ 1-form field $F_{1, A}$ corresponding to 14 defect branes, and  the $\alpha=-4$ 2-forms $F_{2, A_1 ...A_4}$ in the ${\bf 1001}$ and $F_{2, AB}$ in the ${\bf 104}$, giving the domain walls:
  \begin{eqnarray}
& &  F_{2, m_1 \pm ... m_4 \pm }^+ \rightarrow {7 \choose 4} \times 2^4 = 560 \quad ,\nonumber \\
& & F_{2, m\pm m\pm } \rightarrow 14 \quad .
\end{eqnarray}
Finally, the space-filling branes result from the $\alpha=-4$ fields $F_{3, A_1 ...A_7}^+$ in the ${\bf 1716}$ and $F_{3, A, B_1 ...B_4}$ in the ${\bf 11648}$, the $\alpha=-5$ fields $G_{3, A_1 A_2 A_3 \dot{a}}$ in the ${\bf 17472}$ and $G_{3, ABa}$ in the ${\bf 5824}$ and the $\alpha=-6$ field $H_{3, A , B_1 ...B_5 }$ in the ${\bf 24024}$. All these fields belong to the U-duality representation ${\bf 147250}$. Counting the corresponding branes gives
  \begin{eqnarray}
& & F_{3, m_1 \pm ...m_7 \pm}^+ \rightarrow \frac{1}{2} \times {7 \choose 7} \times 2^7 = 64\,, \nonumber\\
& & F_{3, m_1 \pm , m_1 \pm ...m_4 \pm } \rightarrow {7 \choose 4} \times 2^4 \times 4 = 2240\,, \nonumber \\
& & G_{3 , m_1 \pm ...m_3 \pm \dot{a}} \rightarrow { 7 \choose 3} \times 2^3 \times 2^{6-3} = 2240\,, \nonumber \\
& & G_{3, m\pm m\pm a} \rightarrow 7 \times 2 \times 2^{6-1} = 448\,, \nonumber \\
& & H_{3 , m_1 \pm, m_1 \pm ...m_5\pm} \rightarrow {7 \choose 5} \times 2^5 \times 5 = 3360 \quad .
\end{eqnarray}

\begin{table}\scriptsize
\begin{center}
\begin{tabular}{|c|c||c|c|c|c|c|c|c|}
\hline \rule[-1mm]{0mm}{6mm} brane & U repr & $\alpha=0$   & $\alpha=-1$ &
$\alpha=-2$ & $\alpha=-3$ & $\alpha=-4$ & $\alpha=-5$ & $\alpha=-6$ \\
\hline
\hline \rule[-1mm]{0mm}{6mm}
0-brane & $240\subset {\bf {248}}$  & $14 \subset{\bf 14}$  & $64 \subset {\bf {64}}$
& $84\subset {\bf 91} $ & $64 \subset {\bf \overline{64}} $& $14 \subset {\bf 14}$ & &  \\
\hline \rule[-1mm]{0mm}{6mm}
1-brane & $2160\subset {\bf {3875}}$  & $1\subset {\bf 1}$ & $ 64 \subset {\bf \overline{64}}$ & $280 \subset {\bf 364}$ &
$ 448\subset {\bf 832}
$ & $560 \subset {\bf 1001}$ & $448 \subset \overline{\bf 832}$ & $280 \subset {\bf 364}$ \\
& & & & & & $14 \subset {\bf 104}$ & & \\
\hline \rule[-1mm]{0mm}{6mm}
2-brane & $17280 \subset {\bf {147250}}$  &  & $64 \subset {\bf
{64}}$ & $560 \subset {\bf 1001}$ & $1344 \subset {\bf {4928}}$ & $2240 \subset  {\bf 11648}$ & $2240 \subset {\bf \overline{17472}}$ & $3360 \subset {\bf 24024}$   \\
 & & & & & &  $64 \subset {\bf \overline{1716}}$  & $448 \subset {\bf 5824}$ & \\
\hline
\end{tabular}
\end{center}
  \caption{\sl The  non-standard branes of $D=3$ maximal
supergravity. The U-duality group is $\text{E}_{8(8)}$ and
the T-duality group is $\text{SO}(7,7)$. The branes with $\alpha$ from $-7$ to $-11$ have not been included but can be read from the Table observing that for a given $p$-brane, each $\alpha$ is mapped to $-\alpha -4 (p+1)$ by conjugation. \label{D=3branescan}}
\end{table}

The results are summarised in Table \ref{D=3branescan}. All the lower values of $\alpha$ are automatically determined using S-duality.

\section{Brane Orbits and Charge Constraints}

In the previous sections we have discussed the classification of maximally
supersymmetric branes within representations of T-duality and
U-duality. In this section we want to determine the orbits filled
out by the action of the symmetry group (either T-duality or
U-duality) on each given brane. In \cite {Lu:1997bg} the U-duality
orbits of all standard branes in any dimension were determined. Our
analysis will extend these results, and determine the U-duality
orbits of all the non-standard branes. We will first determine in
all cases the T-duality orbits.

As we will discuss further below, the charges associated to
maximally supersymmetric $p$-branes sit in the highest weight orbit
of the relevant irreducible representation, namely the orbit to
which the highest weight of the representation belongs. This orbit
is always \textit{unique}, it has the largest isotropy subgroup
(\textit{stabiliser}), and thus minimal dimension among all the
possible orbits in the same representation. For such orbits a
general classification is possible based on the identification of
all the roots associated to generators that annihilate the highest
weight. These generators, together with the Cartan generators with
zero weight, form the stabiliser of the orbit. More precisely,
denoting with $G$ a generic group, the highest weight $\Lambda $ is
annihilated by all the generators associated to the positive roots,
together with a subset of the negative roots. The highest weight
also identifies the Cartan generators that stabilise the orbit:
these are the Cartan generators associated to all the vanishing
Dynkin labels of the highest weight, plus all possible linear
combinations of the other Cartan generators that give a vanishing
contribution. If $n$ is the number of non-vanishing Dynkin labels of
the highest weight, the number of such possible linear combinations
is $n-1$. The generators associated to the negative roots that
belong to the stabiliser, together with the positive root
counterpart and with the already mentioned Cartan generators, form
the semi-simple part of the orbit stabiliser that we denote with
$H$. The remaining positive roots determine the non-semi-simple part
of the stabiliser, which decomposes into irreducible representations
of $H$ itself that we collectively denote by $\mathbf{R}$.
Summarising, the orbit reads
\begin{equation}
G/[H\ltimes {T}_{\mathbf{R}}]\quad ,
\end{equation}
where ${T}_{\mathbf{R}}$ are (generalised) translations corresponding to $%
\mathbf{R}$. This is the approach used in \cite{Lu:1997bg} and which
we will adopt in this paper. The whole procedure is discussed in
detail in Appendix A for the particular case of $\text{SO}(6,6)$.
The reader can see how the results can easily be generalised to any
simple Lie group, and in particular to all the simply-laced groups
discussed in this paper. In this section we will summarise the
results of that analysis. One can also perform an analysis of the
embedding of $H$ in $G$ as a result of a chain of maximal embeddings
yielding a 5- (or more extended) grading structure, and also
uniquely relate the orbit to a particular proper sub-manifold of the
scalar manifold of the maximal supergravity theory under
consideration. These issues will be discussed in detail in a
forthcoming publication \cite{toappear}.

As mentioned in the Introduction, we will also determine the
constraints satisfied by the charges. It turns out that these
constraints are always maximal, in the sense that they are always
the maximal number of constraints that one can impose without making
the whole representation vanish. This again corresponds to the fact
that we are dealing with highest weight orbits. For U-duality
orbits, our constraints generalise the ones obtained in
\cite{Ferrara:1997ci,FG-1,Lu:1997bg} for the standard potentials.

In the rest of this section we will determine all the maximally
supersymmetric brane orbits. In subsection 4.1 we will discuss the
T-duality orbits and the corresponding constraints, starting from
the results of the previous sections. A basic example explaining the general procedure behind the orbit construction is worked out in the Appendix. In subsection 4.2 we will discuss the U-duality orbits. We will first
review the results of \cite {Lu:1997bg} for standard branes and, next,  we
will generalise the analysis to non-standard branes.

\subsection{T-duality Orbits}

In this subsection we will determine the T-duality orbits for
all maximally supersymmetric branes in any dimension. We will apply
directly the method discussed in Appendix A to all the fields listed
in Tables \ref {table1} and \ref{table2}. To refer to the various
$\text{SO}(d,d)$ representations, we will sometimes refer to the
Dynkin indices corresponding to the nodes in the Dynkin diagram of
fig. \ref{DdDynkindiagram}. We will also mention the constraints
that the corresponding charges satisfy, and show that these are the
maximal constraints that one can impose without setting to zero the
whole representation. As already mentioned, this implies that the
orbit is unique and of minimal dimension.

\begin{table}\small
\begin{center}
\begin{tabular}{|c|c|c|}
\hline \rule[-1mm]{0mm}{6mm} $\alpha$ & field  & stabiliser\\
\hline
\hline \rule[-1mm]{0mm}{6mm}
$0$ & $B_{1,A}$&     $\text{SO}(d-1,d-1) \ltimes T_{\bf 2(d-1)} $                \\[.1truecm]
& $B_2$ & $\text{SO}(d,d)$ \\[.1truecm]
\hline \rule[-1mm]{0mm}{6mm}
$-1$ & $C_{2n+1,a}\,, C_{2n,{\dot a}}$ & $ \text{SL}(d,\mathbb{R}) \ltimes T_{\bf d \choose 2} $\\[.1truecm]
\hline  \rule[-1mm]{0mm}{6mm}
$-2$ & $ D_{D-4}$ & $\text{SO}(d,d)$  \\[.1truecm]
& $D_{D-3, A}$ &  $\text{SO}(d-1,d-1) \ltimes T_{\bf 2(d-1)}$                \\[.1truecm]
& $D_{D-2, A_1 A_2}$ &  $(\text{SL}(2,\mathbb{R}) \times \text{SO}(d-2,d-2) )\ltimes (T_{\bf (1,1)} \times T_{\bf (2,2(d-2))}) $   \\[.1truecm]
& $D_{D-1, A_1 A_2 A_3}$ &  $(\text{SL}(3,\mathbb{R}) \times \text{SO}(d-3,d-3) )\ltimes (T_{\bf (\overline{3},1)} \times T_{\bf (3,2(d-3))})$   \\[.1truecm]
& $D_{D, A_1 ...A_4} $ &  $(\text{SL}(4,\mathbb{R}) \times \text{SO}(d-4,d-4) )\ltimes (T_{\bf ({6},1)} \times T_{\bf (4,2(d-4))}) $   \\[.1truecm]
\hline  \rule[-1mm]{0mm}{6mm}
$-3$ & $E_{D-2, \dot{a}}$ &   $\text{SL}(d,\mathbb{R}) \ltimes T_{\bf d \choose 2}$\\[.1truecm]
& $E_{D-1, A \dot{a}}$ &          $( \text{SL}(d-1,\mathbb{R})\times \text{SO}(1,1) ) \ltimes ( T_{\bf d-1} \times T_{\bf {d -1 \choose 2}}  \times T_{\bf \overline{d-1}})$                                            \\[.1truecm]
& $E_{D, A_1 A_2 \dot{a}} $ &                    $\big( \text{SL}(2,\mathbb{R}) \times \text{SL}(d-2,\mathbb{R})\times \text{SO}(1,1)\big ) \ltimes ( T_{\bf (2,d-2)} \times T_{\bf (1,{d -2 \choose 2}) }  \times T_{\bf (2,\overline{d-2})} \times T_{\bf (1,1)})$                                             \\[.1truecm]
\hline  \rule[-1mm]{0mm}{6mm}
$-4$ & $F^+_{D, A_1 ...A_d}$ &   $\text{SL}(d,\mathbb{R}) \ltimes T_{\bf d \choose 2}$\\[.1truecm]
\hline
\end{tabular}
\end{center}
  \caption{Stabilisers  of the T-duality orbits $\text{SO}( d,d) / [ H \ltimes T_{\bf R}]
$ for the maximally supersymmetric branes associated to the fields
listed in Table \ref{table1}. For the $D$ fields, the particular cases in which $d  \leq 4$ are discussed in the text. Note that the last $E$ field (and its corresponding orbit) only exists in 8 dimensions and below.
  \label{table1orbits}}
\end{table}

We start our analysis by considering the $B$ fields, that is the
fields with $\alpha=0$. For $B_{1,A}$ one obtains the orbit with stabiliser
\begin{equation}
\text{SO}(d-1,d-1) \ltimes T_{\mathbf{2(d-1)}} \quad ,
\label{stabiliserofalightlikevector}
\end{equation}
where, following the strategy discussed in Appendix A, one obtains the
translation generators $T_{\mathbf{2(d-1)}}$ as the representation with $
\text{SO}(d-1,d-1) $ highest weight $\Lambda_{\alpha_1}$. This is the stabiliser of a lightlike vector of $\text{SO}(d,d)$, which implies that the $\alpha=0$ 0-branes are in T-duality lightcone directions. The field
$B_2$ is a T-duality singlet, and thus has trivial orbit, in the
sense that the stabiliser is the whole $\text{SO}(d,d)$ group.

We next consider the $C$ fields, that belong to the spinor
representations of highest weight  $(0,0,0,...,0,1)$ or
$(0,0,0,...,1,0)$ depending on  whether they have odd or even
rank. One obtains that the $H$ part of the stabiliser is
$\text{SL}(d,\Bbb{R})$, while the remaining
positive roots collect in the representation with highest weight either $%
\Lambda _{\alpha _{d}}$ or $\Lambda _{\alpha _{d-1}}$, which is in
either case the representation of $\text{SL}(d,\Bbb{R})$ with two
antisymmetric indices in the fundamental. The stabiliser is thus
\begin{equation}
\text{SL}(d,\Bbb{R})\ltimes T_{\mathbf{{\binom{d}{2}}}}\quad ,
\end{equation}
namely the ``pure'' spinor orbit, see \textit{e.g.}
\cite {Pol-pure}.

\begin{table}
\begin{center}
\begin{tabular}{|c|c|c|}
\hline \rule[-1mm]{0mm}{6mm} $\alpha$ & field   & stabiliser\\
\hline
\hline \rule[-1mm]{0mm}{6mm}
$-4$ & $F_{D-1,A_1\cdots A_{d-3}}$ &    $(\text{SL}(d-3,\mathbb{R}) \times \text{SO}(3,3) )\ltimes (T_{\bf ({d-3 \choose 2},1)} \times T_{\bf (d-3,6)})$       \\[.1truecm]
& $F_{D,A,B_1\cdots B_{d-3}}$ &    $(\text{SL}(d-4,\mathbb{R}) \times \text{SO}(3,3) \times \text{SO}(1,1))\ltimes (T_{\bf ({d-4 \choose 2},1)} $  \\[.1truecm]
& & $  \times T_{\bf (d-4,6)} \times T_{\bf (1,6)} \times  T_{\bf (d-4,1)} \times T_{\bf (\overline{d-4},1)}  ) $       \\[.1truecm]
 & $D=4~(d=6)$: $F_2$  &  $SO(6,6)$  \\[.1truecm]
& $D=3 ~(d=7)$: $F_{1, A} \quad F_{2, AB}$ &   $\text{SO}(6,6) \ltimes T_{\bf 12}$       \\[.1truecm]
\hline \rule[-1mm]{0mm}{6mm}
$-5$  & $G_{D,A_1\cdots A_{d-4},{\dot a}}$ &                    $( \text{SL}(d-4,\mathbb{R}) \times \text{SL}(4,\mathbb{R})\times \text{SO}(1,1) ) \ltimes ( T_{\bf (d-4,4)}$ \\[.1truecm]
& &  $\times T_{\bf (1,{6}) }  \times T_{\bf ({d-4 \choose 2 },1)}  \times T_{\bf (d-4, \overline{4})}) $                                     \\[.1truecm]
&$D=4~(d=6)$: $G_{3,a} $ &  $\text{SL}(6,\mathbb{R}) \ltimes T_{\bf 15}$  \\[.1truecm]
& $D=3~(d=7)$: $G_{2, Aa} \quad G_{3 ,AB a}$ &   $( \text{SL}(6,\mathbb{R})\times \text{SO}(1,1) ) \ltimes ( T_{\bf 6} \times T_{\bf {6 \choose 2}}  \times T_{\bf \overline{6}}) $                 \\[.1truecm]
\hline \rule[-1mm]{0mm}{6mm}
$-6$ &$D=4~(d=6)$: $H_{4,A_1 A_2 A_3 A_4}$& $(\text{SL}(4,\mathbb{R}) \times \text{SO}(2,2) )\ltimes (T_{\bf ({6},1)} \times T_{\bf (4,4)}) $\\[.1truecm]
&$D=3~(d=7)$: $H_{2,A_1 A_2 A_3 }$ &   $(\text{SL}(3,\mathbb{R}) \times \text{SO}(4,4) )\ltimes (T_{\bf (\overline{3},1)} \times T_{\bf (3,8)}) $   \\[.1truecm]
& $D=3~(d=7)$: $H_{3,A,B_1\cdots B_5}$&
  $(\text{SL}(4,\mathbb{R}) \times \text{SO}(2,2) \times \text{SO}(1,1))\ltimes (T_{\bf ({6},1)} $  \\[.1truecm]
& & $  \times T_{\bf (4,4)} \times T_{\bf (1,4)} \times  T_{\bf (4,1)} \times T_{\bf (\overline{4},1)}  ) $     \\[.1truecm]
\hline
\end{tabular}
\end{center}
  \caption{Stabilisers  of the T-duality orbits $\text{SO}( d,d) /[H \ltimes T_{\bf R}]
$ for the maximally supersymmetric branes associated to the fields
listed in Table \ref{table2}.
  \label{table2orbits}}
\end{table}

One can perform the same analysis for all the branes resulting from
the fields listed in Table \ref{table1}, that are all the branes
that arise from 10 dimensions by means of suitable wrapping rules.
The resulting stabilisers are listed in Table \ref{table1orbits}. A
special attention is required for the $D$ fields, for which one can
perform the analysis of Appendix A in a straightforward way if
$d\geq 5$ (that means in five dimensions and below), obtaining the
orbits listed in the Table. If $d=4$, that is in six dimensions, the
$D_{D}$ field splits into a selfdual and anti-selfdual representation
(carrying either a vector or a tensor worldvolume multiplet). For
each representation one obtains the stabiliser
\begin{equation}
\text{SL}(4,\Bbb{R})\ltimes T_{\mathbf{6}}\quad ,
\end{equation}
which is the stabiliser given in the Table substituting $d=4$. If
$d=3$, that is in seven dimensions, the $D_{D}$ orbit disappears
(which is consistent with the fact that the field does not give rise
to branes \cite {arXiv:1102.0934}), while the $D_{D-1}$ orbit splits
into a selfdual and anti-selfdual part, for each of which the stabiliser is
given by the $D_{D-1}$ line of Table \ref{table1orbits} with $d=3$.
The same pattern applies to eight and nine dimensions. Similarly,
the last of the $E$ fields (and its corresponding orbit) only exists
in 8 dimensions and below, which is consistent with the fact that
the orbit in Table \ref{table1orbits} is
meaningless if $d=1$. One can also see from the Table that, for instance, the $%
\alpha =-4$ branes and the $\alpha =-1$ branes have the same orbit.
This is because the representations have the same Dynkin indices
different from zero.

We now proceed with deriving the orbits of the branes associated to
the fields listed in Table \ref{table2}. As already anticipated in
this paper, for all these branes there is no straightforward
ten-dimensional origin. The derivation of the stabiliser of the
orbit again follows the strategy discussed in Appendix A. We simply list here
the results in Table \ref {table2orbits}. Unless
otherwise specified, the fields in the Table are in generic
dimension $D=10-d$.

We now comment on the constraints that the corresponding charges
satisfy. We first consider the branes whose corresponding fields
belong to tensor representations of T-duality, that is the $\alpha
$-even branes. We have shown in Section 2 (see also
\cite{arXiv:1102.0934} for the $\alpha =-2$ case) that the branes
correspond to the T-duality components of the fields that are
identified by the $\text{SO}(d,d)$ lightlike directions $m\pm n\pm
...$ such that for antisymmetrised indices one has $m\neq n$, while
any non-antisymmetrised index has to be parallel to any of the
antisymmetrised ones. This means that if the charge contains an
index $n+$, it will not contain the index $n-$. This is precisely
the opposite property of the $\text{SO} (d,d)$ metric  $\eta
_{n+n-}$ in lightcone coordinates which is only non-zero if one
index is $n+$ and the other index $n-$. This implies that the charge
constraints  can be neatly summarised by the conditions
\begin{equation}
QQ\eta =0\,,
\end{equation}
regardless of which pair of indices is contracted. For the case of
the vector representation of T-duality, this condition is the
standard lightlike vector condition $Q_{A}Q_{B}\eta ^{AB}=0$. This
condition identifies the orbit with stabiliser given in
Eq.~\eqref{stabiliserofalightlikevector}. For all the other
representations this invariant condition puts a covariant quantity
to zero, and this is in all cases the highest possible amount of
constraints one can impose without setting to zero the whole
representation. This condition identifies the highest weight orbits
with even $\alpha $ that we have derived in this subsection.

The case of $\alpha $ odd, in which  the fields belong to
tensor-spinor representations, is similar, in the sense that in all
cases all the constraints that can be set to zero non-trivially are
actually set to zero. Let us discuss the spinor representations in
more detail. In six dimensions, that is $d=4$, the charges satisfy
the condition
\begin{equation}
Q^{a}Q^{b}C_{ab}=0\,,\qquad Q^{\dot{a}}Q^{\dot{b}}C_{\dot{a}\dot{b}}=0\,,
\end{equation}
because the charge conjugation matrix is symmetric. Similarly, in
five dimensions, that is $d=5$, the charges satisfy the pure spinor
constraint\,\footnote{We refer to the
Appendix of \cite{arXiv:1102.0934} for the details on the properties
of the Gamma matrices of $\text{SO}(d,d)$.}
\begin{equation}
Q^{a}Q^{b}(C\Gamma _{A})_{ab}=0\,,\qquad Q^{\dot{a}}Q^{\dot{b}}(C\Gamma _{A})_{%
\dot{a}\dot{b}}=0\quad .
\end{equation}
Similar constraints apply to the other cases.

To conclude, we stress that while for standard branes the
constraints are such that the number of supersymmetric branes always
equals the dimension of the representation, for non-standard branes
these constraints are somehow `stronger' and one always gets fewer
branes than the number of components of the corresponding
representation.

\subsection{\label{U-Duality-Hw}U-duality Orbits}

In this subsection we want to determine the U-duality orbits of all
the maximally supersymmetric branes in any dimension. As in the
previous subsection, we follow the strategy of Appendix A. At the
end of this subsection we will  comment on the constraints that the corresponding
brane charges satisfy.

We start our analysis by shortly reviewing the cases of IIB and
$D=9$. In
this case the simple part of the U-duality symmetry is $\text{SL}(2,\Bbb{R})$%
, which has only one positive root. In classifying the orbits we distinguish between two cases. Either the field is a singlet,
in which case the whole $\text{SL}(2,%
\Bbb{R})$ is a stabiliser and the orbit is trivial, or the field transforms non-trivially under U-duality, in which case the
orbit is given by
\begin{equation}
\text{SL}(2,\Bbb{R})/T_{\mathbf{1}}\quad .
\end{equation}

\begin{table}\small
\begin{center}
\begin{tabular}{|c|c|c|c|}
\hline \rule[-1mm]{0mm}{6mm}   field  &  representation & Dynkin labels & stabiliser\\
\hline
\hline \rule[-1mm]{0mm}{6mm}
 $A_{1, Ma}$ &    ${\bf (\overline{3},2)}$ & $(1,0) \oplus 1 $ &        $\text{GL}(2,\mathbb{R}) \ltimes ( T_{\bf 2} \times T_{\bf 1} )$         \\[.1truecm]
\hline \rule[-1mm]{0mm}{6mm}
 $A_{2}^M$ &    ${\bf ({3},1)}$ & $(0,1) \oplus 0 $ &   $( \text{SL}(2,\mathbb{R}) \ltimes T_{\bf 2} ) \times \text{SL}(2,\mathbb{R})$               \\[.1truecm]
\hline \rule[-1mm]{0mm}{6mm}
 $A_{3, a}$ &    ${\bf ({1},2)}$ & $(0,0) \oplus 1 $ &    $\text{SL}(3,\mathbb{R}) \times T_{\bf 1}$             \\[.1truecm]
\hline \rule[-1mm]{0mm}{6mm}
 $A_{4,M}$ &    ${\bf (\overline{3},1)}$ & $(1,0) \oplus 0 $ &      $( \text{SL}(2,\mathbb{R}) \ltimes T_{\bf 2} ) \times \text{SL}(2,\mathbb{R})$               \\[.1truecm]
\hline \rule[-1mm]{0mm}{6mm}
 $A_{5}^M{}_a$ &    ${\bf ({3},2)}$ & $(0,1) \oplus 1 $ &             $\text{GL}(2,\mathbb{R}) \ltimes ( T_{\bf 2} \times T_{\bf 1} )$           \\[.1truecm]
\hline \rule[-1mm]{0mm}{6mm}
 $A_{6,M}{}^N$ &    ${\bf ({8},1)}$ & $(1,1) \oplus 0 $ &      $(\text{SO}(1,1)\ltimes (T_{\bf 1} \times T_{\bf 1} \times T_{\bf 1}    ) \times \text{SL}(2,\mathbb{R})$        \\[.1truecm]
\hline \rule[-1mm]{0mm}{6mm}
 $A_{6,i}$ &    ${\bf ({1},3)}$ & $(0,0) \oplus 2 $ &        $\text{SL}(3,\mathbb{R}) \times T_{\bf 1}$             \\[.1truecm]
\hline \rule[-1mm]{0mm}{6mm}
 $A_{7}^{MN}{}_a $ &    ${\bf ({6},2)}$ & $(0,2) \oplus 1 $ &     $\text{GL}(2,\mathbb{R}) \ltimes ( T_{\bf 2} \times T_{\bf 1} )$                 \\[.1truecm]
\hline \rule[-1mm]{0mm}{6mm}
 $A_{8}^{MN}{}_P$ &    ${\bf ({15},1)}$ & $(1,2) \oplus 0 $ &       $(\text{SO}(1,1)\ltimes (T_{\bf 1} \times T_{\bf 1} \times T_{\bf 1}    ) \times \text{SL}(2,\mathbb{R})$              \\[.1truecm]
\hline
\end{tabular}
\end{center}
  \caption{The U-duality stabilisers of all supersymmetric branes in eight dimensions.
  The U-duality group is $\text{SL}(3,\Bbb{R}) \times \text{SL}(2,\Bbb{R})$.
  \label{D=8orbits}}
\end{table}

We  now consider all the other cases from eight to three
dimensions. In any dimension the U-duality symmetry $\text{E}_{d+1(d+1)}$
is given by the Dynkin diagram of fig. \ref{Ed+1diagram}.
The fields that we discuss in this
paper, that is the fields associated to maximally supersymmetric branes, belong to $%
\text{E}_{d+1(d+1)}$ representations that are given in all cases by eq. %
\eqref{Udualityreprrealroot}, with as only exceptions  the 1-forms and 8-forms in eight dimensions, as discussed in section 3. We first consider the
eight-dimensional
case. In this case the symmetry is $\text{SL}(3,\Bbb{R}) \times \text{SL}(2,%
\Bbb{R})$, and the Dynkin labels are
\begin{equation}
(n_1 , n_2 ) \oplus n_3
\end{equation}
with $n_1$ and $n_3$ labelling $\text{SL}(3,\Bbb{R})$ highest weights and $%
n_3$ being the $\text{SL}(2,\Bbb{R})$ highest weight. The
representation of
the 1-form is given by the Dynkin labels $(1,0)\oplus 1$,\footnote{
The reason why the 1-forms in 8-dimensions do not follow the general pattern of eq. \eqref{Udualityreprrealroot} is explained in section 3.}
 and following
the analysis of the Appendix one obtains the orbit \cite{Lu:1997bg}
\begin{equation}
[ \text{SL}(3,\Bbb{R}) \times \text{SL}(2,\Bbb{R}) ] / [ \text{GL}(2,\Bbb{R}%
) \ltimes ( T_{\mathbf{2}} \times T_{\mathbf{1}} ) ] \quad .
\end{equation}
All the other representations are given by Eq.
\eqref{Udualityreprrealroot}, with the exception of the 8-form,
which belongs to the $\mathbf{({15},1)}$ and whose Dynkin labels are
$(1,2) \oplus 0 $. Applying the methods of Appendix A we get all the
orbits for all branes. The results are summarised in Table
\ref{D=8orbits}. For the standard branes we recover the results of
\cite{Lu:1997bg}.

\begin{table}\small
\begin{center}
\begin{tabular}{|c|c|c|c|}
\hline \rule[-1mm]{0mm}{6mm}   field  &  representation & Dynkin labels & stabiliser\\
\hline
\hline \rule[-1mm]{0mm}{6mm}
 $A_{1, MN}$ &    ${\bf \overline{10}}$ & $(1,0,0,0)  $ &      $(\text{SL}(3,\mathbb{R})\times \text{SL}(2,\mathbb{R})) \ltimes T_{({\bf  3,2 })}$         \\[.1truecm]
\hline \rule[-1mm]{0mm}{6mm}
 $A_{2}^M$ &    ${\bf {5}}$ & $(0,0,1,0)  $ &         $\text{SL}(4,\mathbb{R}) \ltimes T_{\bf 4}$        \\[.1truecm]
\hline \rule[-1mm]{0mm}{6mm}
 $A_{3, M}$ &    ${\bf \overline{5}}$ & $(0,0,0,1)  $ &             $\text{SL}(4,\mathbb{R}) \ltimes T_{\bf 4}$          \\[.1truecm]
\hline \rule[-1mm]{0mm}{6mm}
 $A_{4}^{MN}$ &    ${\bf {10}}$ & $(0,1,0,0) $ &         $(\text{SL}(3,\mathbb{R})\times \text{SL}(2,\mathbb{R})) \ltimes T_{({\bf  3,2 })}$            \\[.1truecm]
\hline \rule[-1mm]{0mm}{6mm}
 $A_{5, M}{}^N$ &    ${\bf {24}}$ & $(0,0,1,1)  $ &     $(\text{SL}(3,\mathbb{R})\times \text{SO}(1,1) ) \ltimes ( T_{\bf 1 } \times T_{\bf 3} \times T_{\bf \overline{3}})$            \\[.1truecm]
\hline \rule[-1mm]{0mm}{6mm}
 $A_{6}^{M,NP}$ &    ${\bf \overline{40}}$ & $(0,1,1,0)  $ &            $(\text{SL}(3,\mathbb{R})\times \text{SO}(1,1) ) \ltimes ( T_{\bf 1 } \times T_{\bf 3} \times T_{\bf {3}})$          \\[.1truecm]
\hline \rule[-1mm]{0mm}{6mm}
 $A_{6,MN}$ &    ${\bf \overline{15}}$ & $(0,0,0,2)  $ &          $\text{SL}(4,\mathbb{R}) \ltimes T_{\bf 4}$               \\[.1truecm]
\hline \rule[-1mm]{0mm}{6mm}
 $A_{7,M}^{NP} $ &    ${\bf {70}}$ & $(0,0,2,1)  $ &        $(\text{SL}(3,\mathbb{R})\times \text{SO}(1,1) ) \ltimes ( T_{\bf 1 } \times T_{\bf 3} \times T_{\bf \overline{3}})$                 \\[.1truecm]
\hline
\end{tabular}
\end{center}
  \caption{The U-duality stabilisers of all supersymmetric branes in seven dimensions.
  The U-duality group is $\text{SL}(5,\Bbb{R})$. Observe that the convention for the order of the Dynkin labels follows from the Dynkin diagram of fig. \ref{Ed+1diagram} for $d=3$, which is not the one conventionally used for $\text{A}_4$.
  \label{D=7orbits}}
\end{table}

We now consider the seven-dimensional case. We point the attention
of the reader to the fact that the Dynkin labels that one obtains
from the Dynkin diagram of fig. \ref{Ed+1diagram} with $d=3$ are not
the conventional ones for $\text{A}_{4}$. Let us illustrate the
procedure for the $0$-branes, whose highest weight orbit has
$H=\text{SL}(3,\Bbb{R})\times \text{SL}(2,\Bbb{R})$, where the
simple roots of $\text{SL}(3,\Bbb{R})$ are $\alpha _{2},\alpha _{3}
$ while the simple root of $\text{SL}(2,\Bbb{R})$ is $\alpha _{4}$.
In order to determine $T_{\mathbf{R}}$, one observes that the
highest weight
\begin{equation}
\Lambda _{\alpha _{1}}=\alpha _{1}+\alpha _{2}+\alpha _{3}+\alpha
_{4}
\end{equation}
has Dynkin labels $(0,1)\oplus 1$ with respect to $H$. This gives
the stabiliser \cite{Lu:1997bg}
\begin{equation}
(\text{SL}(3,\Bbb{R})\times \text{SL}(2,\Bbb{R}))\ltimes T_{(\mathbf{3,2}%
)}\quad .
\end{equation}
Similarly, one can obtain the stabiliser in all the other cases. The
results are summarised in Table \ref{D=7orbits}. The standard-brane
orbits were already obtained in \cite{Lu:1997bg}.

\begin{table}\small
\begin{center}
\begin{tabular}{|c|c|c|c|}
\hline \rule[-1mm]{0mm}{6mm}   field  &  representation & Dynkin labels & stabiliser\\
\hline
\hline \rule[-1mm]{0mm}{6mm}
 $A_{1, \dot{\alpha}}$ &    ${\bf {16}}$ & $(1,0,0,0,0)  $ &        $\text{SL}(5,\mathbb{R}) \ltimes T_{\bf 10}$         \\[.1truecm]
\hline \rule[-1mm]{0mm}{6mm}
 $A_{2, M}$ &    ${\bf {10}}$ & $(0,0,0,1,0)  $ &     $\text{SO}(4,4)\ltimes T_{\bf 8}$            \\[.1truecm]
\hline \rule[-1mm]{0mm}{6mm}
 $A_{3, \alpha}$ &    ${\bf \overline{16}}$ & $(0,0,0,0,1)  $ &          $\text{SL}(5,\mathbb{R}) \ltimes T_{\bf 10}$          \\[.1truecm]
\hline \rule[-1mm]{0mm}{6mm}
 $A_{4,MN}$ &    ${\bf {45}}$ & $(0,0,1,0,0) $ &       $(\text{SL}(2,\mathbb{R})\times \text{SO}(3,3) ) \ltimes (T_{\bf (1,1)} \times T_{\bf (2,6)})$          \\[.1truecm]
\hline \rule[-1mm]{0mm}{6mm}
 $A_{5, M\alpha}$ &    ${\bf {144}}$ & $(0,0,0,1,1)  $ &           $( \text{SL}(4,\mathbb{R}) \times \text{SO}(1,1) ) \ltimes (T_{\bf 6} \times T_{\bf 4} \times T_{\bf \overline{4}})$        \\[.1truecm]
\hline \rule[-1mm]{0mm}{6mm}
 $A_{6,M,NP}$ &    ${\bf {320}}$ & $(0,0,1,1,0)  $ &          $( \text{SO}(3,3) \times \text{SO}(1,1) ) \ltimes (T_{\bf 6} \times   T_{\bf 6} \times T_{\bf 1} \times T_{\bf 1})$       \\[.1truecm]
\hline \rule[-1mm]{0mm}{6mm}
 $A_{6,M_1 ...M_5}^+$ &    ${\bf \overline{126}}$ & $(0,0,0,0,2)  $ &          $\text{SL}(5,\mathbb{R}) \ltimes T_{\bf 10}$           \\[.1truecm]
\hline
\end{tabular}
\end{center}
  \caption{The U-duality stabilisers of all supersymmetric branes in six dimensions.
The U-duality group is $\text{SO}(5,5)$. Observe that the convention for the order of the Dynkin labels follows from the Dynkin diagram of fig. \ref{Ed+1diagram} for $d=4$, which is not the one conventionally used for $\text{D}_5$.
  \label{D=6orbits}}
\end{table}

The Dynkin diagram of the six-dimensional U-duality symmetry
$\text{SO}(5,5)$ arises from fixing $d=4$ in fig. \ref{Ed+1diagram}.
As in the seven-dimensional case, we emphasise that the
convention for Dynkin labels that results is not the conventional
one. The analysis of the orbits in this
case in very similar to the one performed in detail in Appendix A for $\text{%
SO}(6,6)$. Here we just mention the particular case of the
6-form, in the irreducible mixed-symmetry representation with three
vector indices. One has
\begin{equation}
H = \text{SO}(3,3) \times \text{SO}(1,1) \quad ,
\end{equation}
and the remaining positive roots collect in representations
determined by the highest weights
\begin{eqnarray}
& & \Lambda_{2\alpha_3 + \alpha_4 } \rightarrow \mathbf{1} \quad , \nonumber \\
& & \Lambda_{\alpha_3 + \alpha_4} \rightarrow \mathbf{6} \quad , \nonumber \\
& & \Lambda_{\alpha_3 } \rightarrow \mathbf{6}\quad ,  \nonumber \\
& & \Lambda_{\alpha_4} \rightarrow \mathbf{1} \quad .
\end{eqnarray}
The corresponding stabilisers of this case and all the others are
listed in Table \ref{D=6orbits}. The orbits corresponding to the
standard potentials, i.e.~the 1-forms, 2-forms and 3-forms, were determined in \cite{Lu:1997bg}.

We now consider the five-dimensional case. Using the method of
Appendix A, the orbits can be determined in all cases, and the
results are summarised in Table \ref{D=5orbits}. As an example, we
 consider the 3-forms. The highest weight corresponds to the
Dynkin labels $(0,0,0,0,0,1)$, which leads to
$H=\text{SL}(6,\Bbb{R})$ with simple roots $\alpha _{1},\alpha
_{2},\alpha _{3},\alpha _{4},\alpha _{5}$. The other positive roots
collect in representations of highest weights
\begin{eqnarray}
&&\Lambda _{2\alpha _{6}}\rightarrow \mathbf{1} \quad , \nonumber \\
&&\Lambda _{\alpha _{6}}\rightarrow \mathbf{20}\quad .
\end{eqnarray}
The stabiliser is thus
\begin{equation}
\text{SL}(6,\Bbb{R})\ltimes T_{\mathbf{1}}\times
T_{\mathbf{20}}\quad ,
\end{equation}
and similarly one can obtain all the other orbits. The complete set of
results is summarised in Table \ref{D=5orbits}.

\begin{table}\small
\begin{center}
\begin{tabular}{|c|c|c|c|}
\hline \rule[-1mm]{0mm}{6mm}   field  &  representation & Dynkin labels & stabiliser\\
\hline
\hline \rule[-1mm]{0mm}{6mm}
 $A_{1, M}$ &    ${\bf {27}}$ & $(1,0,0,0,0,0)  $ &     $\text{SO}(5,5) \ltimes T_{\bf 16}$             \\[.1truecm]
\hline \rule[-1mm]{0mm}{6mm}
 $A_{2}^{ M}$ &    ${\bf \overline{27}}$ & $(0,0,0,0,1,0)  $ &    $\text{SO}(5,5) \ltimes T_{\bf 16}$              \\[.1truecm]
\hline \rule[-1mm]{0mm}{6mm}
 $A_{3}{}^{ \alpha}$ &    ${\bf {78}}$ & $(0,0,0,0,0,1)  $ &             $\text{SL}(6,\mathbb{R}) \ltimes ( T_{\bf 20} \times T_{\bf 1})$          \\[.1truecm]
\hline \rule[-1mm]{0mm}{6mm}
 $A_{4}{}^{MN}$ &    ${\bf {351}}$ & $(0,0,0,1,0,0) $ &       $(\text{SL}(2,\mathbb{R}) \times \text{SL}(5,\mathbb{R})) \ltimes ( T_{\bf (1,5)} \times T_{\bf (2,\overline{10})})$           \\[.1truecm]
\hline \rule[-1mm]{0mm}{6mm}
 $A_{5}^{ M\alpha}$ &    ${\bf \overline{1728}}$ & $(0,0,0,0,1,1)  $ &      $(\text{SL}(5,\mathbb{R}) \times \text{SO}(1,1) ) \ltimes ( T_{\bf 1} \times T_{\bf \overline{10}} \times T_{\bf 10} \times T_{\bf 5})$            \\[.1truecm]
\hline
\end{tabular}
\end{center}
  \caption{The U-duality stabilisers of all supersymmetric branes in five dimensions.
  The U-duality group is $E_{6(6)}$.
  \label{D=5orbits}}
\end{table}

The same technique can be applied to four and three dimensions, and
the results are listed in Tables \ref{D=4orbits} and
\ref{D=3orbits}. As a final example we consider the
$\mathbf{147250}$ of $\text{E}_{8(8)}$, which are the 3-forms in
three dimensions. In this case $H=\text{SL}(8,\Bbb{R})$ generated by
the simple roots $\alpha_1 ...\alpha_7$, and the remaining positive
root collect in representations of highest weights
\begin{eqnarray}
& & \Lambda_{3\alpha_8 } \rightarrow \mathbf{\overline{8}}\quad ,  \nonumber \\
& & \Lambda_{2\alpha_8} \rightarrow \mathbf{28} \quad , \nonumber \\
& & \Lambda_{\alpha_8 } \rightarrow \mathbf{\overline{56}} \quad .
\end{eqnarray}
The corresponding stabiliser is given in the third row of Table \ref
{D=3orbits}.

We now comment on the U-duality constraints on the charges.
First of all, the constaints turn out to be always quadratic in the
charges. We  consider as a general example the constraint on the
defect branes. These branes are always in the adjoint of the
U-duality group $G$, and their number is always
$\mathrm{dim}\,G-\mathrm{rank}\,G$ \cite
{arXiv:1109.2025,Bergshoeff:2011se}. If we denote with $\alpha $ the
index in the adjoint of $G$, we write the constraint on the charge
$Q^{\alpha }$ as the ``adjoint light-cone''
\begin{equation}
Q^{\alpha }Q^{\beta }g_{\alpha \beta }=0\,,
\end{equation}
where $g_{\alpha \beta }$ is the Cartan-Killing metric. So the
charges are in the lightlike directions in adjoint space. Given that
the group $G$ is always maximally non-compact, the number of
timelike (that is non-compact) directions is
$1/2\,[\mathrm{dim}\,G+\mathrm{rank}\,G]$ while the number of
spacelike (that is compact) directions is $1/2\,[\mathrm{dim}\,G-\mathrm{rank}\,G]$%
. This implies that the number of lightlike directions is $\mathrm{dim}\,G-%
\mathrm{rank}\,G$, which is indeed equal to the number of defect
branes. In this case, the highest weight orbit is the
\textit{minimal nilpotent} orbit.\,\footnote{For the theory of
nilpotent orbits of simple Lie algebras, see
\textit{e.g.}~\cite{CMcG-book}; recent physical applications can be
found in, e.g., \cite{Bergshoeff:2008be,BN-1,BMP-1}.} Furthermore,
the relation between maximally supersymmetric branes and highest
weight orbits which we establish generalises the link between
maximally supersymmetric black holes ($0$-branes) and
minimal nilpotent (adjoint) orbits, noticed \textit{e.g.} in
\cite{BN-1}.

\begin{table}\small
\begin{center}
\begin{tabular}{|c|c|c|c|}
\hline \rule[-1mm]{0mm}{6mm}   field  &  representation & Dynkin labels & stabiliser\\
\hline
\hline \rule[-1mm]{0mm}{6mm}
 $A_{1, M}$ &    ${\bf {56}}$ & $(1,0,0,0,0,0,0)  $ &          $\text{E}_{6(6)}\ltimes T_{\bf 27}$       \\[.1truecm]
\hline \rule[-1mm]{0mm}{6mm}
 $A_{2}^{ \alpha}$ &    ${\bf {133}}$ & $(0,0,0,0,0,1,0)  $ &    $ \text{SO}(6,6) \ltimes (T_{\bf 1} \times T_{\bf 32})$             \\[.1truecm]
\hline \rule[-1mm]{0mm}{6mm}
 $A_{3,A}$ &    ${\bf {912}}$ & $(0,0,0,0,0,0,1)  $ &        $\text{SL}(7,\mathbb{R}) \ltimes ( T_{\bf 7} \times T_{\bf \overline{35}})$         \\[.1truecm]
\hline \rule[-1mm]{0mm}{6mm}
 $A_{4, \alpha\beta}$ &    ${\bf {8645}}$ & $(0,0,0,0,1,0,0) $ &   $(\text{SL}(6,\mathbb{R}) \times \text{SL}(2,\mathbb{R}) ) \ltimes ( T_{\bf (1,2)} \times T_{\bf (15,1)}\times T_{\bf (\overline{15},2 )})$               \\[.1truecm]
\hline
\end{tabular}
\end{center}
  \caption{The U-duality stabilisers of all supersymmetric branes in four dimensions.
  The U-duality group is $E_{7(7)}$.
  \label{D=4orbits}}
\end{table}

In IIB and in nine dimensions, denoting schematically with $Q$ the
charge, in all cases every contraction of $QQ$ with a single
$\epsilon^{\alpha\beta}$ of $\text{SL}(2,\Bbb{R})$ gives zero. This
follows from the fact that all charges are obtained as products of
the fundamental doublet charge $q^\alpha$ \cite{Bergshoeff:2006gs}.
This result generalises to all dimensions, and in all cases the
contraction of $QQ$ with a single invariant tensor gives a vanishing
result. For instance, the 0-branes in eight dimensions have charges
$q^{Ma}$ satisfying the constraint
\begin{equation}
q^{Ma} q^{Nb} \epsilon_{MNP}\epsilon_{ab} =0 \quad .
\end{equation}
Similarly, the 6-branes have charges satisfying
\begin{equation}
Q_{MNa} Q_{PQb} \epsilon^{MPR} =0 \quad .
\end{equation}
Similar constraints can be derived for the other eight-dimensional
branes, and can be seen as resulting from the way the charges are
decomposed in terms of the fundamental (0-brane) charges
\cite{arXiv:1109.1725}.

A similar analysis can be performed in lower dimensions. For
instance, the 0-branes satisfy the constraint
\begin{equation}
Q^{MN}Q^{PQ}\epsilon _{MNPQR}=0  \label{constraint0branesD=7}
\end{equation}
in seven dimensions, the ``pure'' spinor constraints
\begin{equation}
Q^{\dot{\alpha}}Q^{\dot{\beta}}(C\Gamma
_{M})_{\dot{\alpha}\dot{\beta}}=0 \label{constraint0branesD=6}
\end{equation}
in six dimensions and the constraints
\begin{equation}
Q^{M}Q^{N}d_{MNP}=0  \label{constraint0branesD=5}
\end{equation}
in five dimensions \cite{Ferrara:1997ci}. Similar constraints can be
obtained in all the other cases in any dimension.

\begin{table}\small
\begin{center}
\begin{tabular}{|c|c|c|c|}
\hline \rule[-1mm]{0mm}{6mm}   field  &  representation & Dynkin labels & stabiliser\\
\hline \hline \rule[-1mm]{0mm}{6mm}
 $A_{1, \alpha}$ &    ${\bf {248}}$ & $(1,0,0,0,0,0,0,0)  $ &           $\text{E}_{7(7)} \ltimes (T_{\bf 1 } \times T_{\bf 56})$      \\[.1truecm]
\hline \rule[-1mm]{0mm}{6mm}
 $A_{2, M}$ &    ${\bf {3875}}$ & $(0,0,0,0,0,0,1,0)  $ &       $\text{SO}(7,7) \ltimes ( T_{\bf 14} \times T_{\bf 64})$          \\[.1truecm]
\hline \rule[-1mm]{0mm}{6mm}
 $A_{3,A}$ &    ${\bf {147250}}$ & $(0,0,0,0,0,0,0,1)  $ &   $\text{SL}(8,\mathbb{R}) \ltimes ( T_{\bf \overline{8}} \times T_{\bf 28}  \times T_{\bf \overline{56}})$             \\[.1truecm]
\hline
\end{tabular}
\end{center}
  \caption{The U-duality stabilisers of all supersymmetric branes in three dimensions.
  The U-duality group is $E_{8(8)}$.
  \label{D=3orbits}}
\end{table}

Similar to the T-duality constraints discussed in the previous
subsection, the U-duality constraints have the property that  that while for standard branes
the constraints are such that the number of supersymmetric branes
always equals the dimension of the representation, for non-standard
branes these constraints are somehow `stronger' and one always gets
fewer branes than the number of components of the corresponding
representation. We leave this interesting issue for  future
investigation.

Finally, we like to mention that the T-duality
constraints of the previous subsection follow directly from the U-duality
constraints once the branching of the U-duality representations with
respect to T-duality is
performed. For instance, consider the constraint of eq. %
\eqref{constraint0branesD=7} for the 0-branes in seven dimensions.
Under a
T-duality decomposition, the $\mathbf{\overline{10}}$ decomposes according to  $\mathbf{\overline{10}}=\mathbf{6}+%
\mathbf{\overline{4}}$. The $\mathbf{6}$ corresponds to the F0 charge $Q^{ab}$, while
the $\mathbf{\overline{4}}$ corresponds to the D0 charge $Q^{a}$, where $a$ is an
$\text{SL}(4,\Bbb{R})\cong \text{SO}(3,3)$ index. The constraint
becomes for $Q^{ab}$
\begin{equation}
Q^{ab}Q^{cd}\epsilon _{abcd}=0\,,
\end{equation}
which is equivalent to
\begin{equation}
Q^{A}Q^{B}\eta _{AB}=0 \label{lightconeD=7}
\end{equation}
in $\text{SO}(3,3)$ notation. Similarly, from the U-duality constraints %
\eqref{constraint0branesD=6} in six dimensions, using $\mathbf{16}=\mathbf{8}%
_{\mathrm{C}}\oplus \mathbf{8}_{\mathrm{S}}$, we obtain the  $\text{SO}%
(4,4)$ T-duality constraints
\begin{equation}
Q^{\dot{a}}Q^{\dot{b}}C_{\dot{a}\dot{b}}=0,
\end{equation}
and
\begin{equation}
Q^{a}Q^{b}C_{ab}=0\,,
\end{equation}
which selects 8 lightlike direction in both spinor spaces.

It is interesting to see how the T-duality orbits are embedded into the U-duality ones. Let us consider the seven-dimensional example just discussed. The constraint \eqref{constraint0branesD=7} gives a seven-dimensional U-duality orbit inside the ten-dimensional space which is the vector space of the ${\bf {10}}$ (the charges are in representations that are conjugate to the ones to which the  fields belong). Under T-duality, the  ${\bf {10}}$ decomposes in ${\bf 6 \oplus 4}$, giving the charges $Q^A$ and $Q^a$. On the 6-dimensional vector space, and at the origin of the four-dimenisonal one, that is taking $Q^a=0$, the orbit becomes the T-duality orbit which is the lightlike vector orbit of $\text{SO}(3,3)$, which is five-dimensional.  Consistently, the constraint becomes the lighcone condition \eqref{lightconeD=7} as already discussed. Similarly, on the 4-dimensional vector space and at the origin of the six-dimensional one, that is imposing $Q^A =0$, the orbit becomes the four-dimensional T-duality orbit $\text{SO}(3,3) / [ \text{SL}(3,\mathbb{R}) \ltimes {\bf 3}]$, which spans the whole vector space. Consistently, the constraint \eqref{constraint0branesD=7} leads to no constraint when projected on the ${\bf 4}$. While the U-duality vector space is the sum of the T-duality vector spaces, clearly the same is not true for the orbits. It is straightforward  to check that the dimensions do not match. One simple way of understanding this is that one cannot freely span independently the six-dimensional and the four-dimensional spaces due to the further constraint
  \begin{equation}
  Q^A Q^a (C\Gamma_A )_{ab} =0\,,
\end{equation}
which derives from eq.  \eqref{constraint0branesD=7} taking one charge on the ${\bf 6}$ and one charge on the ${\bf 4}$.

\section{\label{Conclusion}Conclusions and Outlook}

In the first part of this work we completed the classification of half-supersymmetric branes
in maximal supergravity theories which we started in our earlier work \cite{Bergshoeff:2010xc,arXiv:1102.0934,arXiv:1108.5067}.
As a general pattern we find that the reduction of the ten-dimensional IIA/IIB branes follow specific wrapping rules.
This includes the reduction of the space-filling branes whose wrapping rule we deduced in this work. As far as we can tell, the T-duality representations of branes which do not contain any wrapped IIA/IIB brane do not obey similar wrapping rules.

In the second part of this work we derived the T-duality and
U-duality orbits to which the half-supersymmetric branes belong. We
found that these orbits are always the highest weight orbits of the
relevant irreducible representation, both for T-duality and
U-duality groups. This relation generalises the link between
supersymmetric black holes ($0$-branes) and minimal
nilpotent (adjoint) orbits, as noticed, \textit{e.g.}, in \cite{BN-1}.
Concerning the U-duality orbits, our results generalise to
non-standard potentials the ones obtained in \cite{Lu:1997bg}.
Similar to \cite{Lu:1997bg}, we find that in general both the
T-duality and U-duality  orbits  do not cover the full
representation spaces. For standard branes, the whole stratification
of the corresponding representation space can be obtained by
including multi-charge solutions with less supersymmetry
\cite{Lu:1997bg,FG-1,ICL-1}. For non-standard branes the situation
is more subtle, since a proper definition of a non-standard brane
configuration requires the inclusion of multi-brane configurations,
with different charges, as well as orientifolds. This issue requires
a further investigation.

We also studied the invariant sets of constraints characterising the
highest weight charge orbits. We found that these are the maximum
number of constraints one can impose without setting the whole
representation to zero. This is in agreement with the absence of a
stratification of the highest weight orbit, and with the fact that
it has the largest stabiliser among all the orbits in which the
relevant representation space stratifies under the action of the
T-duality or U-duality group. In this respect, our results
generalise to non-standard potentials the ones obtained in
\cite{Ferrara:1997ci}.

Our work suggests several interesting further investigations, which
we hope to present in a future publication \cite{toappear}, and
which we here shortly summarise. First of all, in $D=3$, $4$ and $5$
dimensions an interesting feature occurs, namely the fact that in
some cases ($\alpha=-4$ space-filling branes in $D=5$ and $D=4$, as
well as $\alpha=-4$ domain walls and $\alpha=-4$ and $\alpha=-5$
space-filling branes in $D=3$) the relevant U-duality representation
splits into two highest weight orbits related to two different
irreducible representations of T-duality (among the various obtained
from the branching of the  U-duality irreducible representation).
This leads to branes of the same type, that is branes supporting a
worldvolume vector multiplet, belonging to different T-duality
representations for a given $\alpha$. This phenomenon can be
physically understood at least in $D=4$ and $D=5$ as follows. In
dimensions higher than 5, one has 5-branes that are always of two
different types, one supporting a vector multiplet and one
supporting a tensor multiplet, belonging to two different U-duality
representations. When these branes are wrapped, they all give rise to vector multiplet branes and from the U-duality perspective the representations are unified in a single one. Considering in particular the $\alpha=-4$ branes,  we see by looking at Tables \ref{table1} and \ref{table2} that the  space-filling vector branes in $D \geq 6$ come from the field $F_{D,A_1 ...A_d}^+$, while in $D=6$ there is also a space-filling tensor brane coming from the field $F_{6,AB}$. In five and four dimensions, the corresponding U-duality representations unify in a single one, while the T-duality representations remain different.

Secondly, a more detailed study of the stabiliser of the
highest weight orbits shows that it results from (chains of) maximal
embedding(s) yielding a 5- (or higher extended) grading structure and, furthermore, that   to each charge orbit  one
can associate
a proper sub-manifold of the
scalar manifold of the corresponding maximal supergravity theory.
For asymptotically flat branes, such a sub-manifold can be regarded
as the
``moduli space'' of the ADM mass \cite{ADM} of the brane itself (see, \textit{%
e.g.}, \cite{Ferrara-Marrani-2,CFMZ1-D=5,ICL-1}). It would be
interesting to see whether there exists a similar physical interpretation in the case of the non-standard branes.

It is also interesting to consider in more detail the relations with
the theory of standard asymptotically flat branes. They correspond
to brane solutions of maximal supergravity, and in some cases they
can be related to Jordan and Freudenthal triple systems \cite{FG-1},
thus allowing for a covariant approach to the classification of the
orbits based on the theory of invariant polynomials of the
corresponding U-duality representation (see, \textit{e.g.},
\cite{ICL-1} and references therein). In other cases, these branes
exhibit intriguing connections with quantum information theory
\cite{Duff} (see also \cite{BH-qubit,Duff-brane,Levay}).

Last, but not least, it would be interesting to generalise our methods
to non-maximally supersymmetric (or
non-supersymmetric) branes in maximal (or non-maximal) supergravity
theories in various  dimensions.  Similarly, one may generalise our methods from tori to more general manifolds
and try to formulate the whole procedure in terms of the geometric data of these manifolds. All these efforts will be relevant in our search for phenomenological applications of  branes.

\vskip 1.5cm

\section*{Acknowledgements}

We are grateful to Axel Kleinschmidt for discussions on the classification of supersymmetric branes. One of us (A.M.) 
 would like to thank Guillaume Bossard, Paul Sorba, Raymond
Stora and Bert Van Geemen for enlightening discussions.

\vskip 1.5cm

\appendix

\section{Highest Weight Orbits}

In this Appendix we show how the highest weight orbits can be
determined directly by looking at the Dynkin labels of the highest
weight and at the positive roots of the simple group $G$, using the
method that was used in \cite{Lu:1997bg} to determine the highest
weight U-duality orbits of the standard branes.

\begin{figure}[h]
\begin{center}
\begin{picture}(180,70)
\multiput(10,10)(40,0){3}{\circle{10}}
\multiput(130,10)(40,0){2}{\circle{10}}
\put(15,10){\line(1,0){30}}
\multiput(55,10)(40,0){3}{\line(1,0){30}} \put(130,50){\circle{10}}
\put(130,15){\line(0,1){30}} \put(8,-8){$1$}
\put(48,-8){$2$}
 \put(88,-8){$3$}
\put(128,-8){$4$}
\put(168,-8){$5$}  \put(140,47){$6$}
\end{picture}
\caption{\sl The $\text{D}_6$ Dynkin diagram. \label{D6diagram}}
\end{center}
\end{figure}
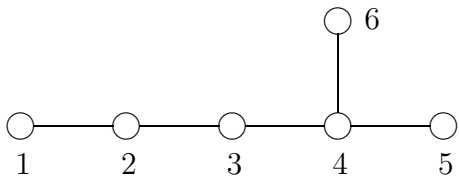

We will focus on the particular example of $\text{SO}(6,6)$, whose
Dynkin diagram is shown in fig. \ref{D6diagram}, but our
considerations are general and apply to all the cases analysed in
Section 4. All the groups we consider are maximally non-compact,
which means that when we write $\text{D}_{6}$ we mean
$\text{SO}(6,6)$, and similarly for all the other cases.

We denote with $\alpha_i$, $i=1,...,6$, the simple roots associated
to the nodes in the Dynkin diagram. The group $\text{D}_6$ has
dimension 66, rank 6
and 30 positive roots. It is useful to list all the positive roots of $\text{%
D}_6$ as the sum of simple roots, and the corresponding height $h$,
which is the number of simple roots that occur in each positive
root. This is summarised in Table \ref{positiverootsD6}. We see from
the Table that the highest value of $h$ is 9.

\begin{table}\small
\begin{center}
\begin{tabular}{|c|c|}
\hline \rule[-1mm]{0mm}{6mm} $h$ & positive root\\
\hline
\hline \rule[-1mm]{0mm}{6mm}
1 & $\alpha_1 \quad \alpha_2 \quad \alpha_3 \quad \alpha_4 \quad \alpha_5 \quad \alpha_6$\\
\hline \rule[-1mm]{0mm}{6mm}
2 & $\alpha_1 + \alpha_2 \quad \alpha_2+\alpha_3 \quad \alpha_3 + \alpha_4 \quad \alpha_4 + \alpha_5\quad \alpha_4 +\alpha_6$\\
\hline \rule[-1mm]{0mm}{6mm}
3 & $ \alpha_1+\alpha_2+\alpha_3 \quad \alpha_2+\alpha_3+\alpha_4 \quad \alpha_3+\alpha_4+\alpha_5\quad \alpha_3 +\alpha_4+\alpha_6 \quad \alpha_4+\alpha_5 +\alpha_6$\\
\hline \rule[-1mm]{0mm}{6mm}
4 & $ \alpha_1+ \alpha_2+ \alpha_3+ \alpha_4 \quad \alpha_2+ \alpha_3+ \alpha_4+ \alpha_5 \quad  \alpha_2+ \alpha_3+ \alpha_4+ \alpha_6\quad  \alpha_3+ \alpha_4+ \alpha_5+ \alpha_6$\\
\hline \rule[-1mm]{0mm}{6mm}
5 & $ \alpha_1 +   \alpha_2+  \alpha_3+ \alpha_4+ \alpha_5 \quad \alpha_1 +   \alpha_2+  \alpha_3+ \alpha_4+ \alpha_6 \quad  \alpha_2+ \alpha_3+ \alpha_4+ \alpha_5+ \alpha_6\quad   \alpha_3+2\alpha_4+\alpha_5+\alpha_6$\\
\hline \rule[-1mm]{0mm}{6mm}
6 & $\alpha_1+\alpha_2+\alpha_3+\alpha_4+\alpha_5+\alpha_6 \quad \alpha_2+\alpha_3+2\alpha_4+\alpha_5+\alpha_6$\\
\hline \rule[-1mm]{0mm}{6mm}
7 & $\alpha_1+\alpha_2+\alpha_3+2\alpha_4+\alpha_5+\alpha_6 \quad \alpha_2+2\alpha_3+2\alpha_4+\alpha_5+\alpha_6$\\
\hline \rule[-1mm]{0mm}{6mm}
8 &  $\alpha_1+\alpha_2+2\alpha_3+2\alpha_4+\alpha_5+\alpha_6$\\
\hline \rule[-1mm]{0mm}{6mm}
9 &  $\alpha_1+2\alpha_2+2\alpha_3+2\alpha_4+\alpha_5+\alpha_6 $\\
\hline
\end{tabular}
\end{center}
  \caption{\sl Table containing the 30 positive roots of $\text{D}_6$ and their height $h$ with respect to  the simple roots.\label{positiverootsD6}}
\end{table}

We can now study the highest weight orbits for various
representations. We denote in general the highest weight with
$\Lambda $, and we identify it in terms of its Dynkin labels
$(n_{1},n_{2},...n_{6})$. As will be clear from the analysis, the
orbits are insensitive to the value of each $n_{i}$, if different
from zero. The stabiliser of the orbit is given by all the Cartan
generators that annihilate the highest weight, together with all the
positive roots (by definition if we sum to $\Lambda $ a positive
root we do not get a weight) and the set of negative roots $-\alpha
$ such that $\Lambda -\alpha $ is not a weight. Denoting with $n$
the number of non-zero entries in the string of Dynkin labels
$(n_{1},n_{2},...n_{6})$, the Cartan generators that stabilise the
orbit are the $6-n$ Cartan generators associated to the vanishing
entries, together with $n-1$ linear combinations of the other $n$
Cartan generators. The negative roots $-\alpha$ , together with the
corresponding positive roots and the $6-n$ Cartan generators, form
the semi-simple part of the stabiliser, while the other $n-1$ Cartan
generators that are stabilisers give rise to additional
$\text{SO}(1,1)$ factors in the stabiliser. We denote with $H$ the
semi-simple part of the stabiliser times the additional
$\text{SO}(1,1)$ factors. What is left are the positive roots whose
corresponding negative roots are not stabilisers. Such roots belong
to a (in general reducible) representation of $H$ that we denote
with $\mathbf{R}$. To summarise, the stabiliser has the form
\begin{equation}
H\ltimes {\ T}_{\mathbf{R}}\quad ,
\end{equation}
where we denote with ${\ T}_{\mathbf{R}}$ the (generalised) translations
corresponding to $\mathbf{R}$.

We start our analysis by considering the representation of highest weight $%
(1,0,0,0,0,0)$, which is the $\mathbf{12}$ of $\text{SO}(6,6)$. One
can show that in this case $H=\text{SO}(5,5)$, generated by the
simple roots $\alpha _{2},\alpha _{3},\alpha _{4},\alpha _{5},\alpha
_{6}$ together with all the
other 15 positive roots in Table \ref{positiverootsD6} that do not contain $%
\alpha _{1}$, the corresponding negative roots and the 5 Cartan
generators. This
leaves  10 more positive roots as stabilisers. To identify the $\text{SO}%
(5,5)$ representation to which these 10 roots belong, we consider
its highest weight, which is the root of Table \ref{D6diagram} with
$h=9$,
\begin{equation}
\Lambda _{\alpha _{1}}=\alpha _{1}+2\alpha _{2}+2\alpha _{3}+2\alpha
_{4}+\alpha _{5}+\alpha _{6}\quad .  \label{10weight}
\end{equation}
We have denoted this weight with an $\alpha _{1}$ label because the
number of times the simple root $\alpha _{1}$ occurs clearly cannot
change within any $\text{SO}(5,5)$ representation. By considering
the scalar product of the weight of Eq.~\eqref{10weight} with the
simple roots $\alpha _{2},\alpha _{3},\alpha _{4},\alpha _{5},\alpha
_{6}$ of $\text{SO}(5,5)$, one obtains the string of Dynkin labels
$(1,0,0,0,0)$, which identifies the highest weight of the
$\mathbf{10}$ of $\text{SO}(5,5)$. This can easily be confirmed by
noticing that all the roots in Table \ref{positiverootsD6}
containing $\alpha _{1}$ form the weights of the $\mathbf{10}$ when
projected on the subspace spanned by the simple roots $\alpha
_{2},\alpha _{3},\alpha _{4},\alpha _{5},\alpha _{6}$. To summarise,
the stabiliser of the highest weight orbit  of the $\mathbf{10}$ of
SO$\left( 6,6\right) $ is given by
\begin{equation}
\text{SO}(5,5)\ltimes T_{\mathbf{10}}\quad .
\end{equation}
This is the stabiliser of a lightlike vector of $\text{SO}(6,6)$,
which implies that the highest weight orbit is the lightcone orbit.

We next consider the highest weight $(0,1,0,0,0,0)$ of the
$\mathbf{66}$, that is the adjoint of $\text{SO}(6,6)$. In this case
one gets
\begin{equation}
H=\text{SL}(2,\Bbb{R})\times \text{SO}(4,4)\quad ,
\end{equation}
where the simple root of ${\text{S}L}(2,\Bbb{R})$ is $\alpha _{1}$
while the simple roots of $\text{SO}(4,4)$ are $\alpha _{3},\alpha
_{4},\alpha _{5},\alpha _{6}$ and all the other positive roots of
$\text{SO}(4,4)$ are those not containing $\alpha _{2}$ in Table
\ref{positiverootsD6}. One is
left with 17 positive roots, which one collects in weights of $\text{SL}(2,%
\Bbb{R})\times \text{SO}(4,4)$ according to the number of times the
$\alpha _{2}$ simple root occurs. At $h=9$ one has
\begin{equation}
\Lambda _{2\alpha _{2}}=\alpha _{1}+2\alpha _{2}+2\alpha
_{3}+2\alpha _{4}+\alpha _{5}+\alpha _{6}\quad ,
\label{sinlgetweight}
\end{equation}
which is a singlet because there are no other roots containing
$2\alpha _{2}$ (and consistently one can show that the scalar
product of $\Lambda _{2\alpha _{2}}$ with $\alpha _{1}$ and with
$\alpha _{3},\alpha _{4},\alpha _{5},\alpha _{6}$ vanishes). At
$h=8$ one has
\begin{equation}
\Lambda _{\alpha _{2}}=\alpha _{1}+\alpha _{2}+2\alpha _{3}+2\alpha
_{4}+\alpha _{5}+\alpha _{6}\quad ,
\end{equation}
which gives the highest weight 1 of $\text{SL}(2,\Bbb{R})$ and the
highest
weight $(1,0,0,0)$ of $\text{SO}(4,4)$, thus giving the representation $%
\mathbf{(2,8)}$. To summarise, the stabiliser of the highest weight
orbit of the $\mathbf{66}$ of $SO\left( 6,6\right) $ is
\begin{equation}
\big(\text{SL}(2,\Bbb{R})\times \text{SO}(4,4)\big)\ltimes
(T_{\mathbf{(1,1)}}\times T_{\mathbf{(2,8)}})\quad .
\end{equation}
This is the  minimal
nilpotent orbit of the adjoint representation.

We now analyse the highest weight $(0,0,1,0,0,0)$, belonging to the $\mathbf{%
220}$. One gets
\begin{equation}
H=\text{SL}(3,\Bbb{R})\times \text{SO}(3,3)\quad ,  \label{Hfor220}
\end{equation}
where $\alpha _{1},\alpha _{2}$ are the simple roots of
$\text{SL}(3,\Bbb{R})
$ while $\alpha _{4},\alpha _{5},\alpha _{6}$ are the simple roots of $\text{%
SO}(3,3)$. The remaining 21 positive roots collect in the
representation with highest weight
\begin{equation}
\Lambda _{2\alpha _{3}}=\alpha _{1}+2\alpha _{2}+2\alpha
_{3}+2\alpha _{4}+\alpha _{5}+\alpha _{6}  \label{triplettweight}
\end{equation}
at $h=9$ and in the representation with highest weight
\begin{equation}
\Lambda _{\alpha _{3}}=\alpha _{1}+\alpha _{2}+\alpha _{3}+2\alpha
_{4}+\alpha _{5}+\alpha _{6}
\end{equation}
at $h=7$. By studying the scalar product of these highest weights
with the
simple roots of $H$ in Eq.~\eqref{Hfor220} one obtains the Dynkin labels $%
(0,1)\oplus (0,0,0)$ and $(1,0)\oplus (1,0,0)$ respectively,
corresponding to the representations $\mathbf{(\overline{3},1)}$ and
$\mathbf{(3,6)}$. The stabiliser of the highest weight orbit of the
$\mathbf{220}$ of SO$\left( 6,6\right) $ is therefore
\begin{equation}
\big(\text{SL}(3,\Bbb{R})\times \text{SO}(3,3)\big)\ltimes (T_{\mathbf{(\overline{3}%
,1)}}\times T_{\mathbf{(3,6)}})\quad .
\end{equation}

One can perform the same analysis in all the other cases. We now
consider some cases in which two Dynkin labels are different from
zero. We start with $(0,0,0,0,1,1)$, which denotes the
$\mathbf{792}$ (a tensor with 5 antisymmetrised indices). In this
case one gets
\begin{equation}
H=\text{SL}(5,\Bbb{R})\times \text{SO}(1,1)\quad ,
\end{equation}
where $\text{SL}(5,\Bbb{R})$ is generated by the simple roots
$\alpha _{1},\alpha _{2},\alpha _{3},\alpha _{4}$, while
the $\text{SO}(1,1)$ factor  is the additional combination of the two remaining
Cartan generators which gives vanishing weight. One is left with 20
positive roots, collecting in the representations with highest
weights
\begin{eqnarray}
&&\Lambda _{\alpha _{5}+\alpha _{6}}=\alpha _{1}+2\alpha
_{2}+2\alpha
_{3}+2\alpha _{4}+\alpha _{5}+\alpha _{6}\qquad (h=9) \quad , \nonumber \\
&&\Lambda _{\alpha _{5}}=\alpha _{1}+\alpha _{2}+\alpha _{3}+\alpha
_{4}+\alpha _{5}\qquad \qquad \qquad \quad \, (h=5) \quad ,  \nonumber \\
&&\Lambda _{\alpha _{6}}=\alpha _{1}+\alpha _{2}+\alpha _{3}+\alpha
_{4}+\alpha _{6}\qquad \qquad \qquad \quad \, (h=5)\quad .
\label{highestweightsgravitinoSO66}
\end{eqnarray}
Performing the scalar product with the simple roots of $\text{SL}(5,\Bbb{%
R})$ one obtains the Dynkin labels $(0,1,0,0)$, $(1,0,0,0)$ and again $%
(1,0,0,0)$, respectively. To summarise, the stabiliser of the highest weight orbit of $%
\mathbf{792}$ of SO$\left( 6,6\right) $ is given by
\begin{equation}
(\text{SL}(5,\Bbb{R})\times \text{SO}(1,1))\ltimes (T_{\mathbf{10}}\times T_{%
\mathbf{5}}\times T_{\mathbf{5}})\quad .
\end{equation}

We now consider the highest weight $(1,0,0,0,0,1)$, which is the
``gravitino'' $\mathbf{\overline{352}}$ representation. In this case
$H$ is again $\text{SL}(5,\Bbb{R})\times \text{SO}(1,1)$, but this
time the simple roots of $\text{SL}(5,\Bbb{R})$ are $\alpha
_{2},\alpha _{3},\alpha _{4},\alpha _{5}$. Again, there are 20
positive roots left, but this time they collect in the
representations with highest weights
\begin{eqnarray}
&&\Lambda _{\alpha _{1}+\alpha _{6}}=\alpha _{1}+2\alpha
_{2}+2\alpha
_{3}+2\alpha _{4}+\alpha _{5}+\alpha _{6}\qquad (h=9)\quad ,   \nonumber \\
&&\Lambda _{\alpha _{6}}=\alpha _{2}+2\alpha _{3}+2\alpha
_{4}+\alpha
_{5}+\alpha _{6}\qquad \qquad \qquad \, (h=7) \quad ,  \nonumber \\
&&\Lambda _{\alpha _{1}}=\alpha _{1}+\alpha _{2}+\alpha _{3}+\alpha
_{4}+\alpha _{5}\qquad \qquad \qquad \quad \, (h=5)\quad .
\end{eqnarray}
Performing the scalar products with the simple roots $\alpha
_{2},\alpha
_{3},\alpha _{4},\alpha _{5}$ one obtains the Dynkin labels $(1,0,0,0)$, $%
(0,1,0,0)$ and $(0,0,0,1)$, respectively. The corresponding
stabiliser of the highest weight orbit of the ``gravitino''
$\mathbf{\overline{352}}$ representation of SO$\left( 6,6\right) $
therefore reads
\begin{equation}
(\text{SL}(5,\Bbb{R})\times \text{SO}(1,1))\ltimes (T_{\mathbf{5}}\times T_{%
\mathbf{10}}\times T_{\mathbf{\overline{5}}})\quad .
\end{equation}

This analysis can be applied to any representation. To conclude this
Appendix, we consider the highest weight with three different Dynkin labels $%
(1,0,1,0,0,1)$, which is the $\mathbf{43680}$ representation.  In
this case
\begin{equation}
H=\text{SL}(2,\Bbb{R})\times \text{SL}(3,\Bbb{R})\times
\text{SO}(1,1)\times \text{SO}(1,1)\quad ,
\end{equation}
where $\alpha _{2}$ is the simple root of $\text{SL}(2,\Bbb{R})$ while $%
\alpha _{4},\alpha _{5}$ are the simple roots of
$\text{SL}(3,\Bbb{R})$. In this example there are two linear
combinations of the remaining three Cartan generators which give
vanishing weight. There are 26 positive roots left, collecting in
the representations with highest weights
\begin{eqnarray}
&&\Lambda _{\alpha _{1}+2\alpha _{3}+\alpha _{6}}=\alpha
_{1}+2\alpha _{2}+2\alpha _{3}+2\alpha _{4}+\alpha _{5}+\alpha
_{6}\qquad (h=9)\quad ,   \nonumber
\\
&&\Lambda _{2\alpha _{3}+\alpha _{6}}=\alpha _{2}+2\alpha
_{3}+2\alpha
_{4}+\alpha _{5}+\alpha _{6}\qquad \qquad \qquad \, (h=7)\quad ,   \nonumber \\
&&\Lambda _{\alpha _{1}+\alpha _{3}+\alpha _{6}}=\alpha _{1}+\alpha
_{2}+\alpha _{3}+2\alpha _{4}+\alpha _{5}+\alpha _{6}\quad \ \qquad
\, (h=7)\quad ,
\nonumber \\
&&\Lambda _{\alpha _{3}+\alpha _{6}}=\alpha _{2}+\alpha _{3}+2\alpha
_{4}+\alpha _{5}+\alpha _{6}\qquad \qquad \quad \qquad \, (h=6)\quad ,   \nonumber \\
&&\Lambda _{\alpha _{1}+\alpha _{3}}=\alpha _{1}+\alpha _{2}+\alpha
_{3}+\alpha _{4}+\alpha _{5}\qquad \qquad \quad \qquad \ \, (h=5) \quad ,  \nonumber \\
&&\Lambda _{\alpha _{3}}=\alpha _{2}+\alpha _{3}+\alpha _{4}+\alpha
_{5}\qquad \qquad \qquad \qquad \qquad \quad \, (h=4) \quad ,  \nonumber \\
&&\Lambda _{\alpha _{6}}=\alpha _{4}+\alpha _{5}+\alpha _{6}\qquad
\qquad
\qquad \qquad \qquad \quad \qquad \ \, (h=3) \quad ,  \nonumber \\
&&\Lambda _{\alpha _{1}}=\alpha _{1}+\alpha _{2}\qquad \qquad \qquad
\qquad \qquad \qquad \qquad \quad \ \ (h=2)\quad .
\end{eqnarray}
The corresponding stabiliser of the highest weight orbit of the
$\mathbf{43680}$ representation of SO$\left( 6,6\right) $ therefore
reads
\begin{eqnarray}
&&(\text{SL}(2,\Bbb{R})\times \text{SL}(3,\Bbb{R})\times \text{SO}%
(1,1)\times \text{SO}(1,1))\ltimes (T_{\mathbf{(2,1)}}\times T_{\mathbf{(1,1)%
}}\times T_{\mathbf{(1,3)}}  \nonumber \\
&&\quad \qquad \times T_{\mathbf{(2,3)}}\times T_{\mathbf{(1,\overline{3})}%
}\times T_{\mathbf{(2,\overline{3})}}\times T_{\mathbf{(1,\overline{3})}%
}\times T_{\mathbf{(2,1)}})\quad .
\end{eqnarray}

This method can be generalised to other simple groups and it is
applied in section 4 to determine both the U-duality and T-duality
highest weight orbits for all the fields discussed in this paper.

\end{document}